\documentclass[reprint,amsmath,amssymb,aps,prb,superscriptaddress,floatfix,footinbib]{revtex4-2}

\usepackage{chemformula}
\usepackage[a4paper,total={6.8in,10in}]{geometry}
\usepackage{mathtools}
\usepackage{graphicx}
\usepackage{xcolor}
\usepackage{physics}
\usepackage[colorlinks, linkcolor= blue, citecolor = blue, urlcolor=blue]{hyperref}
\usepackage[none]{hyphenat}
\usepackage{bm}
\usepackage{comment}
\usepackage[version=4]{mhchem}
\usepackage{wrapfig}
\usepackage{lmodern}
\usepackage{dcolumn}
\usepackage{natbib}

\usepackage{multirow}
\graphicspath{ {./figures/} }
\def\nn{\nonumber}
\def\bea{\begin{eqnarray}}
\def\eea{\end{eqnarray}}
\def\be{\begin{equation}}
\def\ee{\end{equation}}

\def\kb{{\bm k}}
\def\Sb{{\bm S}}
\def\e{\varepsilon}
\def\m{\mathcal}
\def\bal{\begin{aligned}}
\def\eal{\end{aligned}}

\begin{document}
\title{Intrinsic nonlinear thermal Hall transport of magnons: A Quantum kinetic theory approach}



\author{Harsh Varshney}
\email{hvarshny@iitk.ac.in}

\affiliation{Department of Physics, Indian Institute of Technology, Kanpur-208016, India.}

%
\author{Rohit Mukherjee}
\email{rohitmk@iitk.ac.in}
\affiliation{Department of Physics, Indian Institute of Technology, Kanpur-208016, India.}
\affiliation{Department of Physics, Cornell University, Ithaca, New York 14853, USA}

\thanks{Rohit and Harsh contributed equally to the manuscript and are joint first authors.}
\author{Arijit Kundu}
\email{kundua@iitk.ac.in}
\affiliation{Department of Physics, Indian Institute of Technology, Kanpur-208016, India.}
\author{Amit Agarwal}
\email{amitag@iitk.ac.in}
\affiliation{Department of Physics, Indian Institute of Technology, Kanpur-208016, India.}
\begin{abstract}
We present a systematic study of the nonlinear thermal Hall responses in bosonic systems using the quantum kinetic theory framework. We demonstrate the existence of an intrinsic nonlinear boson thermal current, arising from the quantum metric which is a wavefunction dependent band geometric quantity. In contrast to the nonlinear Drude and nonlinear anomalous Hall contributions, the intrinsic nonlinear thermal conductivity is independent of the scattering timescale.  We demonstrate the dominance of this intrinsic thermal Hall response in topological magnons in a two-dimensional ferromagnetic honeycomb lattice without Dzyaloshinskii-Moriya interaction. Our findings highlight the significance of band geometry induced nonlinear thermal transport and motivate  experimental probe of the intrinsic nonlinear thermal Hall response with implications for quantum magnonics.
\end{abstract}
\maketitle
%
\section{Introduction}
Quantum information processing technologies require fast, compact, and power-efficient quantum devices. A promising approach to achieve these goals is to use magnons as carriers of information \cite{application1,application2,application3,application4,YUAN20221}. Magnons are quasiparticles of spin excitations in magnetically ordered materials that carry both energy and spin angular momentum~\cite{maekawa_2012}. Unlike electrons, magnons are electrically neutral and thus not subject to Joule heating, making them attractive candidates for information processing~\cite{HIROHATA2020166711}. Recent experimental advances have demonstrated the feasibility of using spin-orbit torque to excite terahertz magnons and control their quantum states~\cite{terahertzm,terahertzmt}. However, a deeper understanding of magnon transport is crucial to fully exploit their potential for the emerging field of quantum magnonics which combines quantum optics, spintronics, and quantum information processing~\cite{Kouki_2021,neumann_PRL2022_thermal,YUAN20221}. 

\begin{figure}[t]
\centering
    \includegraphics[width = .97 \linewidth]{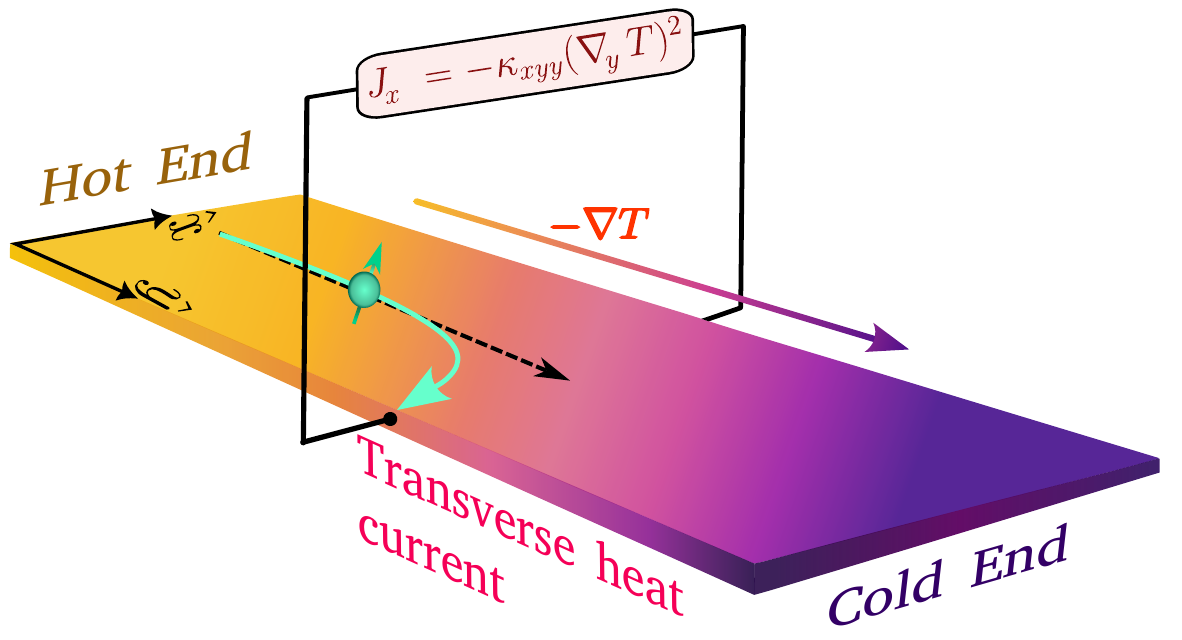}
    \caption{Schematic showing the setup for measuring the nonlinear  
     thermal Hall current of magnons. 
     The temperature gradient ($-\nabla T$) is applied along the $y$ direction and the nonlinear thermal current [$J \propto (\nabla T)^2$] is measured along the $x$ (Hall response) and the $y$ (longitudinal response) direction.}
    \label{fig:schematic}
\end{figure}

The magnon thermal Hall effect is a well-known transport signature of magnons, which arises from the Berry curvature and band topology in bosonic systems~\cite{magnon1,magnon2,magnon3,magnon4,magnonhall5,magnonhall6, owerre2016topological,magnonhall8,magnonhall9,magnonhall10,magnonhall11,matsumoto_PRL2011_theoretical,matsumoto_PRB2011_rotational,xiao_PRB2020_unified,stedman_PRR2020_transport}. Topological magnon systems, such as the magnonic analog of spin Hall insulators~\cite{chenga,chengb,chengc} and magnonic Dirac and Weyl semimetals~\cite{PhysRevB.94.075401,dirac,weyl}, have been extensively explored in recent years. These systems host linear magnon spin Nernst or thermal Hall currents, which are induced by non-collinear spin texture or Dzyaloshinskii-Moriya interaction (DMI). However, in the absence of DMI the linear thermal Hall signal vanishes and transport signatures in the Hall response appear only in the nonlinear response regime. 
This has motivated the recent exploration of exciting  nonlinear transport phenomena in bosonic systems. These include the nonlinear thermal Hall effect~\cite{mukherjee2023nonlinear} [see Fig.~\ref{fig:schematic}], nonlinear spin Nernst effect~\cite{kondo}, and nonlinear optical response~\cite{PhysRevB.98.134422,PhysRevB.100.224411}, among others. The study of topological transport in bosonic systems is not limited to magnons and has also been explored in phonons~\cite{phonon1,phonon2,phonon3}, photons~\cite{phonon3}, and other magnetic excitations~\cite{mukherjee2023nonlinear,mukherjee2021schwinger}. However, a systematic exploration of all possible nonlinear responses in bosonic systems is still lacking. 

To address this, we present a quantum kinetic theory \cite{liboff_book2006_kinetic, nagaosa_PRB2020_quantum,harsh_arx2022_quantum,PhysRevLett.129.227401,PhysRevLett.129.227401,PhysRevB.107.165131} based unified  framework for calculating all the linear and non-linear transport coefficients in bosonic systems. 
We reproduce the known linear and non-linear Drude and anomalous Hall responses. Importantly, we predict the existence of an intrinsic nonlinear thermal Hall conductivity in bosonic systems, which is independent of the magnon scattering timescale and has not been previously explored. We show that the intrinsic nonlinear thermal Hall conductivity arises from the quantum metric and it dominates the thermal Hall response in the regime where the linear and non-linear anomalous thermal Hall responses vanish alongside the Berry curvature. We explicitly demonstrate this by calculating the nonlinear thermal Hall transport of magnons in a ferromagnetic honeycomb lattice with and without DMI. Our results provide new insights into band geometry induced nonlinear transport phenomena in bosonic systems. 

Our quantum kinetic approach provides an elegant alternative to the Boltzmann transport theory \cite{matsumoto_PRL2011_theoretical}, enabling us to include  all interband coherence effects without the need for boundary confining potentials. 
We organize the paper as follows: In Section~\ref{formalism1} and Section~\ref{formalism2}, we describe the quantum kinetic theory formalism for calculating the nonlinear thermal current of bosons, including both intrinsic and extrinsic contributions. Then, in Section~\ref{ferromagnet}, we apply this framework to study magnons in a two-dimensional hexagonal honeycomb lattice. Specifically, we demonstrate the dominance of the intrinsic nonlinear thermal Hall conductivity in the absence of DMI in Sec.~\ref{results and discussion}. 
We discuss some potential materials for observing the intrinsic nonlinear Hall current in Sec.~\ref{Discussions}, and summarize our findings in Sec.~\ref{Conclusion}. 

\section{Nonlinear density matrix with thermal driving for Bosons}\label{formalism1}
To calculate the linear and the nonlinear thermal currents, we first solve  for the density matrix up to second order in the applied temperature gradient $\nabla T$. We use the quantum Liouville equation to calculate the density matrix in the crystal momentum representation $\rho(\kb,t)$. 
The quantum kinetic equation of the density matrix in the presence of a temperature gradient has the form~\cite{nagaosa_PRB2020_quantum, culcer_PRB2017_interband, liboff_book2006_kinetic}
\be \label{eq:qkt_eq}
\frac{\partial \rho(\kb,t)}{\partial t} +\frac{i}{\hbar}\left[ \m{H}_0,\rho(\kb,t)\right] 
= D_T [\rho(\kb,t)]~. 
\ee
Here, ${\m H}_0$ represents the grand canonical Hamiltonian satisfying $\m{H}_0 \vert u^n_{\kb}\rangle = \e^n_{\kb} \vert u^n_{\kb}\rangle$. The energy of the $n$-th band is $\e^n_\kb$ with the corresponding energy eigenstate $\vert u^n_{\kb} \rangle$. 
In Eq.~\eqref{eq:qkt_eq}, 
$[A,B]$ is the commutator bracket for operators $A$ and $B$. For brevity, we use $\rho$ for $\rho(\kb,t)$ and $\e_n$ for $\e^n_{\kb}$ in the rest of the  manuscript. The term on the right-hand side of Eq.~\eqref{eq:qkt_eq} is the thermal driving term~\cite{nagaosa_PRB2020_quantum}, 
\be \label{TD}
D_T (\rho)= - \frac{1}{2\hbar}\bm{E}_T\cdot \left[ \lbrace \m{H}_0 , \pdv{\rho}{\kb} \rbrace - i[\bm{\m{R}_{\kb}},\lbrace \m{H}_0 , \rho \rbrace] \right]~. 
\ee 
Here, $\bm{E}_T \equiv -\bm{\nabla}T/ T  $ is the thermal field~\cite{tatara_PRL2015_thermal} (or temperature gradient) with $T$ being the temperature and the bracket $\lbrace A, B \rbrace$ represents the anticommutator of $A$ and $B$ operators. In Eq.~\eqref{TD}, ${\bm{\m R}}_{\kb}$ is the momentum space Berry connection. In the band-resolve form, it is defined as ${\bm {\m R}}_{np} = i\langle u^n_{\kb} \vert \partial_{\kb} \vert u^p_{\kb} \rangle$.

We use a perturbative approach to solve for the  density matrix in Eq.~\eqref{eq:qkt_eq}. It can be done by expanding $\rho$ in powers of the temperature gradient, $\rho= \rho^{(0)}+ \rho^{(1)} + \rho^{(2)} + \cdots $ where $\rho^{(N)} \propto |\nabla T|^N$ with $N$ being an integer. The equilibrium density matrix is given by $\rho^{(0)} = \sum_n \vert u^n_\kb \rangle \langle u^n_\kb \vert f^B_n$, where $f^B_n \equiv f^B_n(\e_n) = (1 - e^{\e_n/k_B T})^{-1}$ is the Bose-Einstein distribution function. 
The effect of disorder in this framework can be included using the adiabatic switching on~\cite{harsh_arx2022_quantum} of the temperature gradient, $\nabla T \to \nabla T e^{- \eta|t|}$. This approach is equivalent to a relaxation time approximation with $\eta = 1/\tau$ denoting the inverse of the scattering timescale ($\tau$), which is assumed to be a constant for simplicity. 
In the band basis representation, the $N$-th order density matrix elements can be calculated using the following equation\cite{harsh_arx2022_quantum}, 
\be \label{eq:dm_eq}
\frac{\partial \rho_{np}^{(N)}}{\partial t} +\frac{i}{\hbar}\left[ \m{H}_0,\rho^{(N)}\right]_{np}+ \frac{\rho^{(N)}_{np}}{\tau/N} = \left[ D_T (\rho^{(N-1)})\right]_{np}.  
\ee
Here, the subscript $np$ denotes the matrix element of an the operator ${\m O}$ between the $n$-th and the $p$-th energy eigenstates, or ${\m O}_{np} = \bra{u^n_\kb}{\m O}\ket{u^p_\kb}$. 
For the steady-state solution, we ignore the time derivative term of the density matrix. The matrix elements of the commutator term can be evaluated to be   
$\left[ \m{H}_0,\rho^{(N)}\right]_{np} = (\e_n - \e_p) \rho^{(N)}_{np}$. Using this in Eq.~\eqref{eq:dm_eq}, we obtain a recursive solution for the $N$-th order density matrix element, 
\be\label{eq:n-th_dm}  
\rho^{(N)}_{np} = -i\hbar g^{np}_N \left[ D_T (\rho^{(N-1)})\right]_{np}~.
\ee  
Here, we have defined $g^{np}_N = [\e_{np} - i \hbar N/\tau]^{-1}$ 
with $\e_{np} = (\e_n - \e_p)$ being the interband energy gap at a given $\kb$. 
\subsection{First-order density matrix}
%
From Eq.~\eqref{eq:n-th_dm}, we can extract the first-order density matrix elements to be $\rho^{(1)}_{np} = -i\hbar g^{np}_1 \left[ D_T (\rho^{(0)})\right]_{np}$. For the equilibrium density matrix of bosonic excitations (such as  magnons), we have $\rho^{(0)}_{np} = f_n^B \delta_{np}$. Using this, we can easily evaluate  $\left[ D_T (\rho^{(0)})\right]_{np} = -\frac{1}{\hbar} {\bm E}_T \cdot \left[ \e_n \partial_{\kb} f_n^B + i{\bm {\m R}}_{np}(\e_n f_n^B - \e_p f_p^B) \right]$. It will be useful to express the density matrices as a sum of diagonal and off-diagonal parts, $\rho^{(1)} = \rho^{\rm d} + \rho^{\rm o}$. Here, $\rho^{\rm d}$ includes the intra-band processes, while $\rho^{\rm o}$ arises from  the inter-band processes. 
We calculate the diagonal and off-diagonal components of the first-order density matrix to be 
\be  
\bal  
\rho^{\rm d}_{nn} &= -\frac{\tau}{\hbar} \e_n \partial_c f^B_n E^c_T~, \\
\rho^{\rm o}_{np} &= - {\m R}^c_{np} g^{np}_1(\e_n f^B_n - \e_p f^B_p) E^c_T~.
\eal  
\ee  
For brevity, we have defined $\partial_c \equiv \partial_{k_c}$, and used the Einstein summation convention for repeated spatial indices.   
%
\subsection{Second-order density matrix}\label{sec:so_dm}
%
As the second-order density matrix is calculated using the first-order density matrix, both $\rho^{\rm d}$ and $\rho^{\rm o}$ contribute to the diagonal and off-diagonal parts of the second-order density matrix. Therefore, we can express the second-order density matrix as a sum of four terms, $i.e., \  \rho^{(2)} = \rho^{\rm dd} + \rho^{\rm do} + \rho^{\rm od} + \rho^{\rm oo}$. Here, the first two (last two) terms constitute the diagonal (off-diagonal) parts of the second-order density matrix. 
In $\rho^{\alpha \beta}$, the first superscript $\alpha$ (second superscript $\beta$) denotes the diagonal or off-diagonal element of $\rho^{(2)}$. The second superscript $\beta$ in $\rho^{\alpha \beta}$ captures the diagonal or off-diagonal element of $\rho^{(1)}$ from which it originates. 
From Eq.~\eqref{eq:n-th_dm},  we have $\rho^{(2)}_{np} = -i\hbar g^{np}_2 \left[ D_T (\rho^{(1)})\right]_{np}$. The detailed calculation of the different terms of the second-order density matrix is presented in  Appendix~\ref{app:cal_of_so_dm}. We obtain the `dd' part of $\rho^{(2)}$ to be 
\be\label{eq:rho_dd} 
\rho^{\rm{dd}}_{nn} =  \frac{\tau^2}{2\hbar^2} \left[ \hbar \e_n v^n_{b} \partial_c f^B_n  +  \e^2_n \partial_b \partial_c f^B_n \right] E^b_T E^c_T~,
\ee 
where, $v^b_n = \hbar^{-1} \partial_b \e_n $ is the group velocity of the magnons in the $n$-th band along the spatial direction $b = x,~y,~z$. We calculate the `do' component of $\rho^{(2)}$ to be 
\bea 
\rho^{\rm{do}}_{nn} &&= \frac{i\tau}{4\hbar}\sum_{p} (\e_n + \e_p) \left(g^{np}_1\m{R}_{np}^{c}\m{R}_{pn}^{b} + g^{pn}_1 \m{R}_{np}^{b}\m{R}_{pn}^{c}\right) \nn  \\ 
&& \times~  (\e_n f^B_{n}- \e_p f^B_{p} ) E^b_T E^c_T~.
\eea 
The `od' part of the second-order density matrix is given by 
\be
\rho^{\rm{od}}_{np} = \frac{\tau}{\hbar}g^{np}_2 \m{R}_{np}^{b}\left(\e_n ^2 \partial_{c}f^B_{n}-\e_p^2 \partial_{c}f^B_{p} \right)  E^b_T E^c_T~.
\ee
Similarly, we obtain the $\rho^{\rm oo}$ part to be 
\be \label{eq:rho_oo}
\bal  
\rho^{\rm{oo}}_{np}  &= -\frac{i}{2}g^{np}_2 (\e_n + \e_p ) \m{D}^{b}_{np}\left(g_{np}\m{R}_{np}^{c}\xi_{np}\right) E^b_T E^c_T \\ 
&+  \frac{1}{ 2}g^{np}_2 \sum_{q\neq n \neq p} \left[ g^{nq}_1 \m{R}^{c}_{nq}\m{R}^{b}_{qp}(\e_n +\e_q )\xi_{nq}\right. \\ 
&- \left. g^{qp}_1 \m{R}^{b}_{nq}\m{R}^{c}_{qp}(\e_q +  \e_p )\xi_{qp}\right]E^b_T E^c_T~.
\eal 
\ee
Here, we have defined $\xi_{np} \equiv \e_n f^B_n - \e_p f^B_p$ and used the covariant derivative ${\m D}^b_{np} = \partial_b - i({\m R}^b_{nn} - {\m R}^b_{pp})$. The second term of Eq.~\eqref{eq:rho_oo} only contributes in systems having three or more bands. This completes our derivation of the second-order density matrix for a thermal perturbation to a multi-band system. We use it to calculate the nonlinear thermal Hall current in the next section.
%
\begin{figure}[t]
    \centering
    \includegraphics[width=\linewidth]{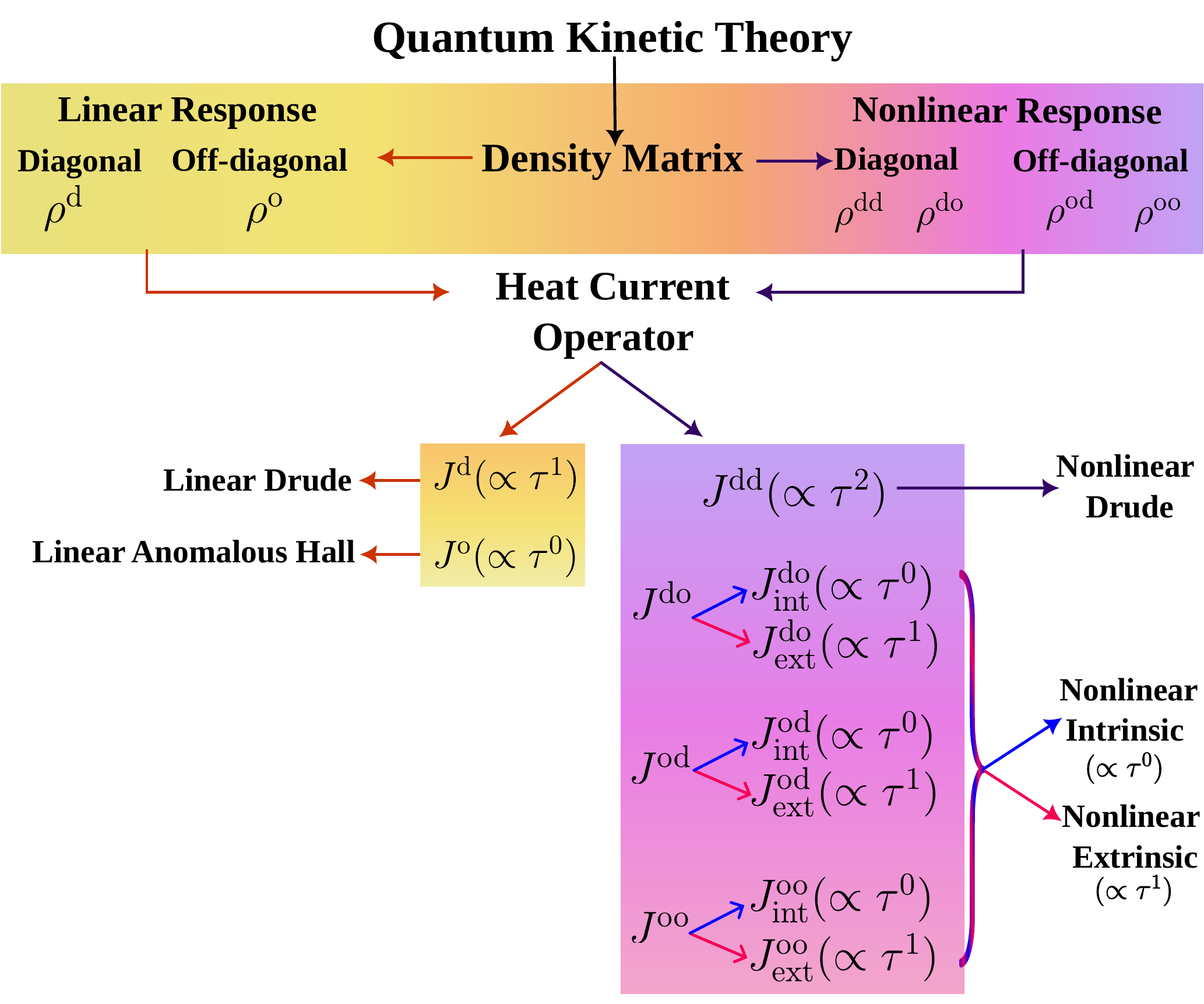}
    \caption{ Origin of the different conductivity contributions in the quantum kinetic theory framework and their scattering time dependence. The off-diagonal terms of the density matrix capture the impact of inter-band coherence. All band geometry induced transport responses arise from the off-diagonal components of the linear or nonlinear density matrix. 
    \label{fig:Tree}}
\end{figure}
\section{Nonlinear Thermal current for BOSONS}\label{formalism2}
%
For bosonic systems, the physically measurable thermal current is identical to the energy current $J^{\rm E}$. 
It turns out that $J^{\rm E}$ can also have a circulating component (curl of energy magnetization for our case), which cannot be measured in transport experiments \cite{qian_nu_PRL2006_berry,xiao_PRB2020_unified}. 
Thus, we have the measurable transport current $J = J^{\rm E} - J^{{\rm E},{\rm mag}}$. The energy current is simply the Brillouin zone sum of the product of the energy current operator (or energy velocity, $\{{\cal H}_0,{\bm v}\}/2$) and the distribution function. 
The energy magnetization current for bosons is specified by $J^{{\rm E},{\rm mag}} =  \nabla_{\bm r} \times {{\bm M}^{\rm E}}$, where ${{\bm M}^{\rm E}}$ is the thermal energy magnetization density in equilibrium. 
The equilibrium energy magnetization density (for a given band $n$) is specified by~\cite{xiao_PRB2020_unified} the sum of two contributions, ${\bm M}^{\rm E}_n = {\bm M}^{{\rm E}}_{{\bm m}_n} + {\bm M}^{\rm E}_{{\bm \Omega}_n}$. 
The first contribution arises from the particle's magnetic moment ${\bm m}_n (\kb)$, and the other arises from its Berry curvature ${\bm \Omega}_n(\kb)$. Explicitly, we have 
%
%
%
%
\be   \label{Mag}
{\bm M}^{\rm E}_n = \int[d\kb] \left(\e_n f^B_0(\e_n) {\bm m}_{n} + \frac{{\bm \Omega}_n}{\hbar} \int^{\infty}_{\e_n} d\e~\e f^B_0(\e)~\right)~.
\ee  
Here, 
$[d\kb] \equiv d^d\kb/(2\pi)^d$ is the integration measure for a $d$-dimensional system.  We note that the 
orbital magnetic moment of the quasiparticle arises from the self-rotation of the wave-packet around its center of mass. The first term in the energy magnetization can be interpreted as the  thermodynamic average of the orbital moment. However, the center of mass motion of the wavepacket gives it another contribution dependent on the Berry curvature [the second term in Eq.~\eqref{Mag}].  

Using the above definition, the linear thermal current (within the quantum kinetic theory) can be expressed as \cite{nagaosa_PRB2020_quantum,harsh_arx2022_quantum}
%
%
\bea\label{eq:th_cur_def}
\bm{J}^{(1)} & = &{\rm{Tr}}\bigg[\frac{1}{2}\{\m{H}_0,\bm{v} \}\rho^{(1)}\bigg] + {\rm{Tr}}\left[(\bm{E}_T \times \bm{m}) \m{H}_0 \rho_0 \right] \nn \\
&&  + 2{\rm Tr}[\bm{E}_T \times \bm{M}_{\bm \Omega}] ~.
\eea 
Here, ${\bm m}$ (${\bm M}_{\bm \Omega}$) is a diagonal matrix with the band resolved quantities ${\bm m}_n$ (${\bm M}_{{\bm \Omega}_n}$) as its diagonal elements expressed in the basis of eigenvectors of ${\cal  H}_0$. 
As a consistency check for Eq.~\eqref{eq:th_cur_def}, we note that it produces all the known linear thermal currents, such as the Drude and the anomalous thermal Hall current. We refer the readers to appendix~\ref{app:linear_th_cur}, where we have presented all the detailed calculations of the linear thermal current.

The nonlinear thermal currents due to the temperature gradient can be obtained similarly. The second-order thermal current is given by\cite{harsh_arx2022_quantum}, 
\be\label{eq:sec_th_cur_def}  
{\bm J}^{(2)} = {\rm Tr}\left[ \frac{1}{2} \{ {\m H}_0, {\bm v}\} \rho^{(2)}\right] + {\rm Tr} \left[ ({\bm E}_T \times {\bm m}) {\m H}_0 \rho^{(1)}\right]~. 
\ee  
The equilibrium magnetization current [the third term on the right-hand side of Eq.~\eqref{eq:th_cur_def}] is linear in the temperature gradient. So, it does not contribute to the second or higher-order thermal current. 
 {A subtle possibility beyond Eq.~\eqref{eq:sec_th_cur_def}, is that the magnetization can get corrections induced by the temperature gradient, $m \to m + m^{T}$. This idea is similar to electric field-induced magnetization, which has been recently explored in Refs.~\cite{xiao_PRB2021_thermoelectric, xiao_PRB2021_adiabatically, kamal_arX2023_nonlinear}. 
The thermal gradient induced magnetization contribution $m^{T}$ can combine with the equilibrium density matrix $\rho^{(0)}$ to possibly have a finite thermal current similar to the second term in Eq.~\eqref{eq:sec_th_cur_def}. However, in this manuscript, we will not explore this possibility; it can be a potential future project.}

Following the four terms of the density matrix, we express the second-order thermal current as  ${\bm J}^{(2)} = {\bm J}^{\rm dd} + {\bm J}^{\rm do} + {\bm J}^{\rm od} + {\bm J}^{\rm oo} + {\bm J}^{\rm mag}$. Here, the first four terms arise from the corresponding component of the second-order density matrix. For example, $\rho^{\rm dd}$ generates ${\bm J}^{\rm dd}$ and so on. The last term ${\bm J}^{\rm mag}$ represents the thermal current arising from the magnetic moment part of Eq.~\eqref{eq:sec_th_cur_def}. 
For the first term in Eq.~\eqref{eq:sec_th_cur_def}, we can show that 
\be \label{eq:j2_part1} 
{\rm Tr}\left[ \frac{1}{2} \{ {\m H}_0, {\bm v}\} \rho^{(2)}\right] = \frac{1}{2}\sum_{n,p,\kb} ({\e}_n + {\e}_p){\bm v}_{pn} \rho^{(2)}_{np}~. 
\ee  
Here, ${\bm v}_{pn}$ denotes the components of the velocity operator defined as $i \hbar {\bm v} = [{\bm r}, {\m H}_0]$. For notational brevity, we have used $\sum_{\kb} \equiv \int [d\kb]$. The elements of the velocity operator are obtained by the covariant derivative of the Hamiltonian (${\bm v}_{pn} = \hbar^{-1} [{\m D}_{\kb}{\m H}_0]_{pn}$) and found to be, 
\be\label{eq:gen_velocity}  
{\bm v}_{pn} = {\bm v}_n \delta_{pn} + i \omega_{pn} {\bm {\m R}}_{pn}~.
\ee  
Here, the first term ${\bm v}_n$ is the group velocity of the quasi-particles in the $n$-th band, while the second term corresponds to the Berry phase correction to the velocity. 

%
%
%
%
%
%

Using these, we calculate the `dd' component of ${\bm J}^{(2)}$ originating from the $\rho^{\rm dd}$ along the  spatial direction $a$ to be, 
\be \label{eq:J_dd}
J^{\rm dd}_{a} =  \frac{\tau^2}{2\hbar^2 } \sum_{n,{\bm k}}  ( \hbar \e_n v^b_n \partial_c f^B_n  +  \e^2_n \partial_{b}\partial_c f^B_n)  \e_n v_n^{a} E^b_T E^c_T~.  
\ee
The $J^{\rm dd}$ component of the thermal current depends only on the group velocity of the quasi-particle. 
Similarly, we calculate the nonlinear thermal current stemming from $\rho^{\rm do}$ to be 
\bea\label{eq:J_do}
J^{\rm{do}}_{a} &=& \frac{i\tau}{2\hbar^2 } \sum_{n,p, {\bm k}}^{p \ne n} g^{np}_1 \m{R}^b_{pn}\m{R}^c_{np} (\e_n + \e_p) (\e_n f^B_{n}- \e_p f^B_{p}) \nn \\ 
&&\times \left(\e_n v_n^{a}- \e_p v_p^{a} \right) E^b_T E^c_T~.
\eea
%
The `od' component of the nonlinear thermal current is given by
\bea\label{J_od} 
 J^{\rm{od}}_{a} & = & \frac{i\tau}{2\hbar^2} \sum_{n,p, \kb}^{p \ne n}\e_{pn} g^{np}_2 ( \e_n + \e_p) \m{R}^a_{pn}\m{R}^b_{np} \nn \\  
&& \times \left(\e_n^2 \partial_{c} f^B_{n} - \e_p^2 \partial_{c} f^B_{p} \right)  E^b_T E^c_T ~.
\eea 
Similarly, we compute the `oo' component of ${\bm J}^{(2)}$ to be 
\begin{widetext}
\be\label{J_oo} 
\bal  
J^{\rm{oo}}_{a} = & - \frac{1}{4\hbar} \sum_{n,p,\kb}^{p \ne n}  \e_{np} g^{np}_2 ( \e_n + \e_p )\m{R}^a_{pn}\bigg[ ( \e_n + \e_p ) \m{D}^b_{np}( g^{np}_1 \m{R}^c_{np} \xi_{np})  
\\
& +  i \sum_q^{q\neq n \neq p} \left( g^{nq}_1\m{R}^{c}_{nq}\m{R}^{b}_{qp}(\e_n +\e_q )\xi_{nq} - 
g^{qp}_1\m{R}^{b}_{nq}\m{R}^{c}_{qp}(\e_q +  \e_p )\xi_{qp} \right)\bigg]E^b_T E^c_T~. 
\eal 
\ee
\end{widetext}
We find that barring the $J^{\rm dd}$ contribution, all the other three contributions depend on the band geometric quantities. 

To calculate the magnetic moment induced $J^{\rm mag}$, we note that the particle magnetic moment of the $n$-th band is given by~\cite{xiao_RMP2010_berry, wang_PRB2022_quantum}
\be\label{eq:pmm}
\bal  
{\bm m}_{n} &= \frac{i}{2\hbar} \langle {\bm \nabla}_\kb u^n_\kb \vert \times [ {\m H}_0 - \e_n] \vert { \bm \nabla}_\kb u^n_\kb \rangle \\ 
&= -\frac{i}{2\hbar} \sum_{p}^{p \ne n} (\e_n - \e_p) ({\bm {\m R}_{np}} \times {\bm {\m R}_{pn}})~.
\eal
\ee 
In writing the second line of the above equation, we have used the completeness relation. Using this, we explicitly calculate $J^{\rm mag}$ and it is given by 
\be\label{eq:mag_cur}  
J_{a}^{\rm{mag}}  = \frac{\tau }{2 \hbar^2} \sum_{n,p,\kb}^{p \ne n}  \e_n^2 (\e_n - \e_p) \Omega^{ab}_{np} \partial_c f^B_n E^b_T E^c_T~.
\ee  
%
We refer the readers to appendix~\ref{app:linear_th_cur} for a detailed calculation of magnetic moment contribution to the linear thermal current. The calculation for the second-order current in  Eq.~\eqref{eq:mag_cur} follows the same approach. 
In in  Eq.~\eqref{eq:mag_cur}, the band geometric quantity $\Omega^{ab}_{np}$ 
is commonly known as the band-resolved Berry curvature. In our calculation, it arises from the band-resolved {quantum geometric tensor~\cite{cheng2013quantum}} defined as $Q^{ab}_{np} = {\m R}^a_{np} {\m R}^b_{pn} \equiv {\m G}^{ab}_{np} -\frac{i}{2} \Omega^{ab}_{np}$, where  ${\m G}^{ab}_{np}$ is the band-resolved quantum metric. Recently quantum metric was measured experimentally~\cite{gianfrate2020measurement} in the context of anomalous Hall drift. Here, we emphasize that the quantum geometric tensor, quantum metric, and Berry curvature are gauge-invariant quantities (see appendix~\ref{gauge invariance} for details). 
The band resolved Berry curvature can also be expressed as $\Omega^{ab}_{np} = -2 \ {\rm Im}[Q^{ab}_{np}] \equiv i ({\m R}^a_{np} {\m R}^b_{pn} - {\m R}^b_{np} {\m R}^a_{pn} )$. In this form, it is explicit that the band-resolved Berry curvature is antisymmetric under the exchange of both spatial and band indices ~\cite{xiao_RMP2010_berry, watanbe_PRX2021_chiral}, or  $\Omega^{ab}_{np} = - \Omega^{ba}_{np}$, and   $\Omega^{ab}_{np} = -\Omega^{ab}_{pn}$. This band-resolved Berry curvature is related to the single band Berry curvature (used in Eq.~\eqref{Mag} for example) via the relation $\Omega^{ab}_n = \sum_p^{p \ne n} \Omega^{ab}_{np}$. 
%

\subsection{Intrinsic and extrinsic contributions to 
thermal Hall currents}
While the previous section calculates all contributions to the nonlinear thermal current of bosons, it is useful to classify them according to their dependence on the scattering time. By doing a systematic expansion in $\tau$, we can express ${\bm J}^{(2)} = {\bm J}^{(2)} (\propto \tau^0) + {\bm J}^{(2)} (\propto \tau^1) + {\bm J}^{(2)} (\propto \tau^2)$. Here, ${\bm J}^{(2)} (\propto \tau^0)$ is the intrinsic part of the nonlinear thermal current independent of the scattering time. 
The other two terms, namely ${\bm J}^{(2)} (\propto \tau^1)$ and ${\bm J}^{(2)} (\propto \tau^2)$ depend on the scattering time and represent the extrinsic nonlinear thermal current. 
This is useful for experiments as the contributions with different $\tau$ dependence can be separated using appropriate scaling laws~\cite{du_NatCom2019_disorder}. This facilitates the   understanding and analysis of the physical mechanism behind the dominant contribution to the thermal current.  This also helps us to identify the dissipationless intrinsic nonlinear thermal currents, which originate from quantum coherence effects and carry the signature of the band geometric quantities. 
Additionally, this helps us to compare results from quantum kinetic theory with the semiclassical thermal transport framework where currents are calculated in orders of the scattering time. 

The different contribution of the nonlinear thermal currents like $J^{\rm do}_a, \ J^{\rm od}_a $, and $J^{\rm oo}_a$ depend upon a $\tau$ dependent function $g^{np}_N$, with $N=1$ or $N=2$. In the dilute impurity limit (DIL) where $\tau \gg 1/\omega_{np}$ or $\tau \omega_{np} \gg 1$, we can  express $g^{np}_N$ as a sum $\tau$-independent and $\tau$-dependent parts, and retain only the dominant $\tau$ contributions (see appendix~\ref{app_C} for details). 
This allows us to express each of the thermal current contributions into an intrinsic (via subscript `int') and extrinsic (via subscript `ext') part as, 
\be\label{eq:th_cur_tau_dependence}
\begin{aligned}
J^{(2)}_a(\propto \tau^0) &= J^{\rm do}_{a,\rm int} + J^{\rm od}_{a,\rm int} + J^{\rm oo}_{a,\rm int}~, \\ 
J^{(2)}_a(\propto \tau^1) &= J^{\rm do}_{a,\rm ext} + J^{\rm od}_{a,\rm ext}+ J^{\rm mag}_{a}~, \\ 
J^{(2)}_a(\propto \tau^2) &= J^{\rm dd}_{a}~.
\end{aligned}
\ee
%
For example, $J^{\rm do}_{a, {\rm int}}$ conveys the intrinsic part of the $J^{\rm do}_a$, while $J^{\rm do}_{a, {\rm ext}}$ corresponds to the extrinsic contribution. 
This is also summarized in Fig.~\ref{fig:Tree}, which explicitly shows the origin of the different thermal Hall contributions in terms of the diagonal and off-diagonal part of the density matrix involved. 

Combining all the intrinsic contributions into one term, we obtain the intrinsic nonlinear thermal current as, 
\bea\label{eq:tot_th_cur_int}   
&& J^{(2)}_a(\propto \tau^0) = \frac{1}{\hbar} \sum_{n,p,\kb}^{p \ne n} \frac{\e_n (\e_n + \e_p) }{  \e_{np}} \times \nn \\ 
&& \bigg[ \frac{f_n^B}{4  \e_{np}} \left\{ (\e^2_n  - \e^2_p) \partial_{a} {\m G}^{bc}_{np} - {\m G}^{bc}_{np} \partial_a  (\e^2_n  - \e^2_p) \right\} \nn \\ 
&& + {\m G}^{ac}_{np} [2\e_n \partial_b f^B_n + f^B_n \partial_b(\e_n + \e_p)]  \bigg]E^b_T E^c_T~.
\eea   
 The intrinsic thermal current explicitly depends on the band-resolved quantum metric ${\m G}^{bc}_{np}$, which is defined as the real part of the quantum geometric tensor, $i.e., \ {\m G}^{bc}_{np} = {\rm Re}[Q^{bc}_{np}] \equiv \frac{1}{2} (\m{R}^b_{np}\m{R}^c_{pn}+\m{R}^c_{np}\m{R}^b_{pn})$. Note that, ${\m G}^{bc}_{np}$ is symmetric under the exchange of both spatial and band indices~\cite{watanbe_PRX2021_chiral, provost_CIMP1980_riemannian, berry_1989_quantum}, $i.e., \ \m{G}^{bc}_{np} = \m{G}^{cb}_{np} ~ {\rm{and}} ~ \m{G}^{bc}_{np} = \m{G}^{bc}_{pn}$. 

Similarly, we obtained the linear $\tau$ dependent nonlinear thermal current through Eq.~\eqref{eq:th_cur_tau_dependence} as 
\be\label{eq:tot_th_cur_ext_1}  
J^{(2)}_a{(\tau^1)} = \frac{\tau}{\hbar^2} \sum_{n,p,\kb}^{p \ne n} \e^3_n \Omega^{ab}_{np} \partial_c f^B_n E^b_T E^c_T~.
\ee  
This contribution is referred to as the \textit{nonlinear anomalous thermal current}. This nonlinear thermal Hall contribution has been calculated earlier using the semiclassical Boltzmann transport in Refs.~[\onlinecite{matsumoto_PRL2011_theoretical, zeng_PRB2020_fundamental, zhou_PRB2022_fundamental, chakraborty_2DM2022_non}]. It is considered as the thermal analog of the Berry curvature dipole induced nonlinear electric Hall effect. 
The quadratic $\tau$-dependent nonlinear thermal current is determined solely by ${\bm J}^{\rm dd}$, and has the same expression as ${\bm J}^{\rm dd}$ in Eq.~\eqref{eq:J_dd}, $i.e. \ J^{(2)}_a (\propto \tau^2) = J^{\rm dd}_{a}$. We refer the readers to appendix~\ref{app_C} for details of this calculation. 
{We note that the intrinsic and Drude nonlinear thermal currents will remain unchanged on going beyond the dilute impurity limit. However, the nonlinear anomalous thermal Hall current gets corrections with different powers of $\tau$. Note that the general expressions for all conductivity contributions (retaining all powers of $\tau$) are presented in Eqs.~\eqref{eq:J_dd}-\eqref{J_oo}. }

The nonlinear thermal current can be expressed as $J_a = - \kappa_{abc} (\tau) \nabla_b T \nabla_c T$ with $\kappa_{abc}(\tau)$ being the nonlinear thermal conductivity tensor of rank three. $\kappa(\tau)$ has all the information of the scattering time dependence of the nonlinear thermal current. 
Under time reversal symmetry, $J_a$ picks up a negative sign, while $\nabla_b T \nabla_c T$ remains unaltered. Therefore, we must have $\kappa_{abc} (-\tau) = - \kappa_{abc} (\tau)$ in the presence of TRS. This works out only if $\kappa_{abc} (\tau)$ depends on the odd powers of $\tau$. Or in other words, in a TRS symmetric system, only the odd power of $\tau$ dependent thermal currents survives. Note that under TRS, we have 
$\e_n(-\kb) = \e_n(\kb) \ {\rm and} \ {\m G}^{ab}_{np} (-\kb) = {\m G}^{ab}_{np} (\kb) $, while $ \ {\bm v}_n(-\kb) = -{\bm v}_n(\kb) $ and $\Omega^{ab}_{np} (-\kb) = -\Omega^{ab}_{np}(\kb)$. Using this, we have explicitly checked that 
%
%
the nonlinear thermal current contributions $J^{(2)}(\propto \tau^2)$ and $J^{(2)}(\propto \tau^0)$ given in Eqs.~\eqref{eq:J_dd} and \eqref{eq:tot_th_cur_int}, respectively vanish in the presence of TRS. 
In contrast, the linear $\tau$ dependent nonlinear thermal current given in Eq.~\eqref{eq:tot_th_cur_ext_1} is finite under TRS. 

Systems where parity and TRS are individually broken, but the composite symmetry of parity-TRS is preserved are also very interesting. In these systems, the Berry curvature is identically zero at each point of the momentum space. This in turn makes the Berry curvature-dependent nonlinear thermal current to vanish. However,  the remaining nonlinear thermal currents, the Drude contribution $J^{(2)}(\tau^2)$ and the intrinsic contribution $J^{(2)}(\tau^0)$ can be finite in such systems. 


Having established the quantum kinetic theory for linear and non-linear transport of bosonic systems, we now explore the thermal Hall current of magnons in two-dimensional hexagonal lattices. 
%
\begin{figure}[t]
    \centering
    \includegraphics[width=\linewidth]{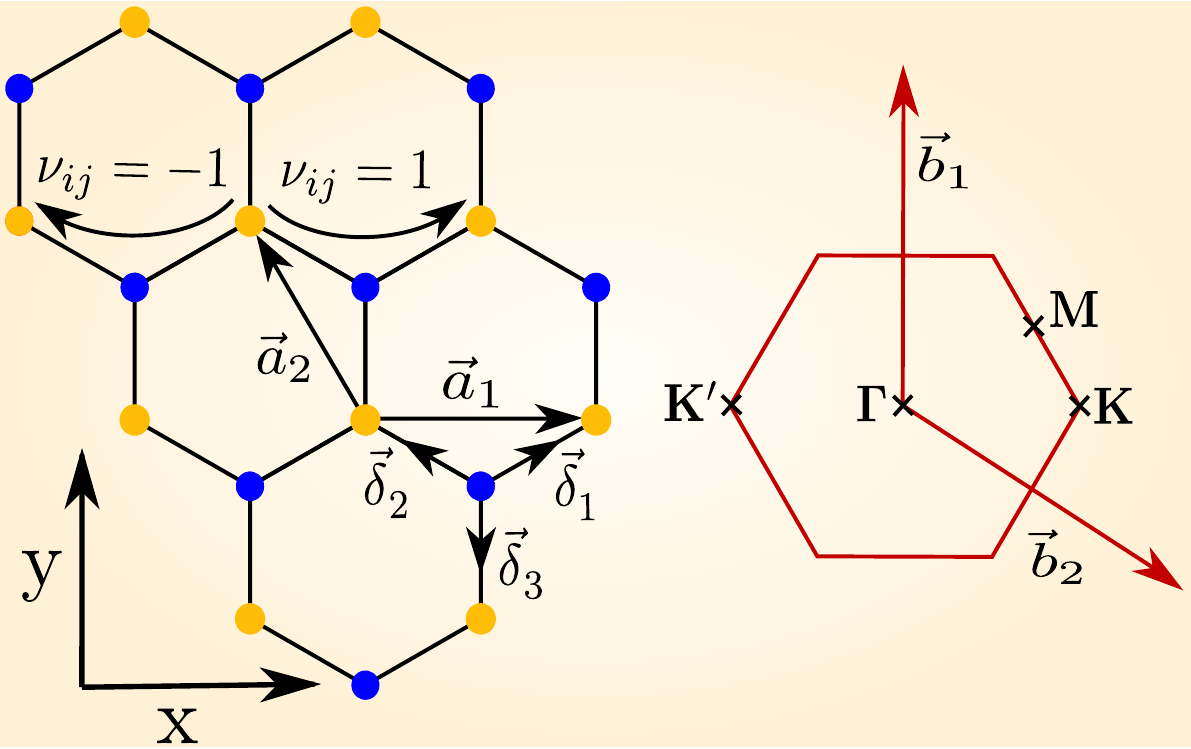}
    \caption{(Left) The unit cell of the honeycomb lattice. $\Vec{a}_{\mu}$ with $\mu = 1,2$ denote the real space lattice vectors. The vectors $\Vec{\delta}_{i}$ with $i = 1,2,3$, connect the nearest neighbors. The sign convention for choosing the Dzyaloshinskii-Moriya interaction is also shown. (Right) The Brillouin zone of the honeycomb lattice with the  reciprocal lattice vectors  $\Vec{b}_{\mu}$, and the high symmetry points marked.}
    \label{fig:Fig3}
\end{figure}
%
\section{Magnons in hexagonal ferromagnets}\label{ferromagnet}
%
%
\begin{figure}[t]
\centering
\includegraphics[width=1.0\linewidth]{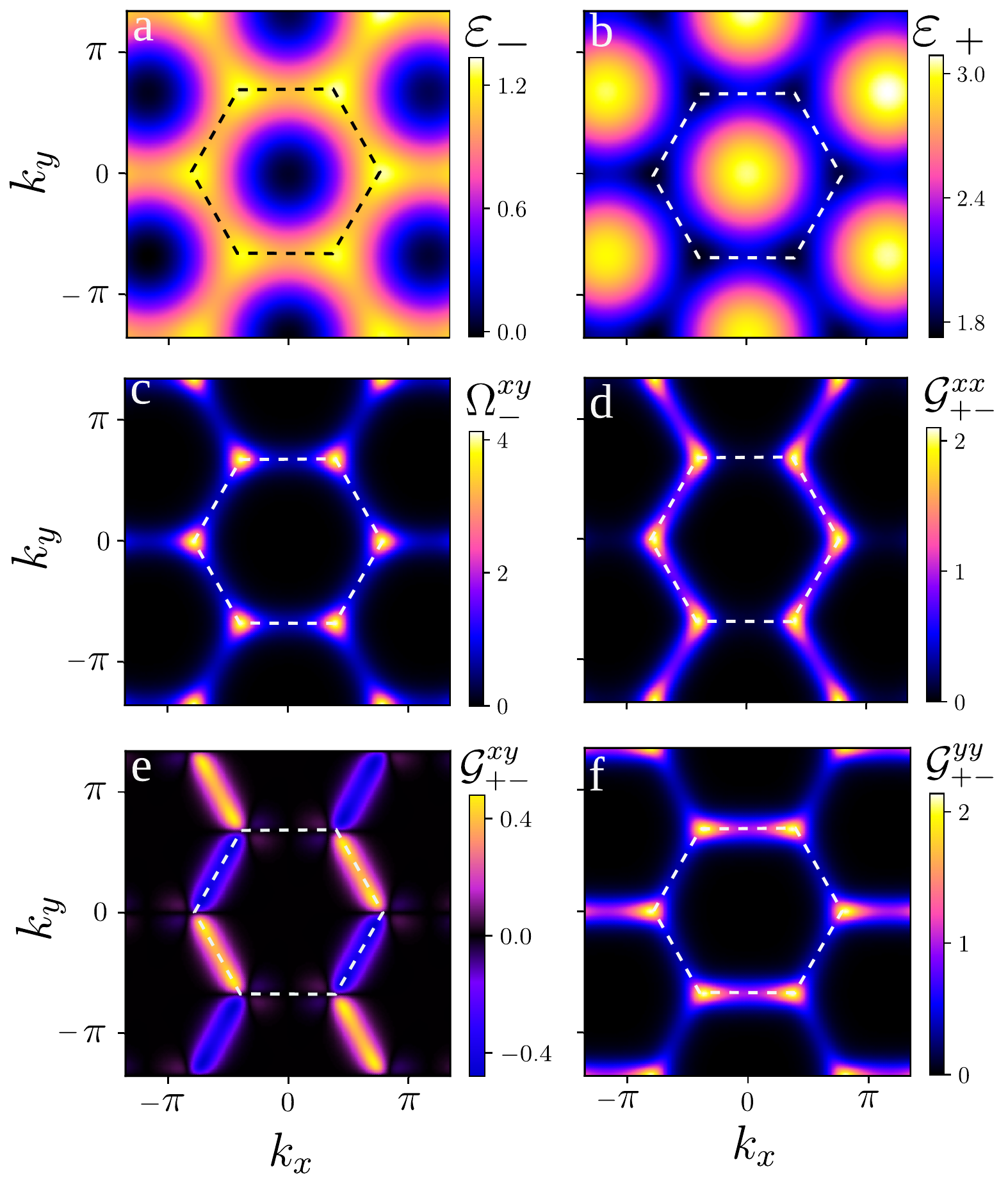}
    \caption{ 
   The color plot of the magnon dispersion in the Brillouin zone for the (a) lower energy band and the (b) higher energy band. Dzyaloshinskii-Moriya interaction opens up a gap at the $K/K'$ point. (c) The distribution of the Berry curvature in momentum-space for the lower energy band. 
    The Berry curvature peaks near the band edges.
    In contrast to the fermionic hexagonal lattices, the Berry curvature of magnon bands in hexagonal ferromagnets retains the same sign in both valleys. 
    (d)-(f) Three distinct components of the quantum metric. We used the following parameters: 
    $\{J_2, D, B, B_s\}  = \{0.01, 0.1,0.001, 0.05\} J_1$ 
    with $J_1 = 1$ meV. In all panels, the hexagon in dashed lines indicates the Brillouin zone. 
    \label{fig:density_plots}}
\end{figure}
%
%
We take a spin model on a single-layer honeycomb lattice as our model to calculate different magnon transport coefficients. Our Hamiltonian consists of different kinds of spin-spin interactions that are relevant for ferromagnetic insulators. The spin Hamiltonian is given by, 
\begin{equation}\label{Hamiltonian_model}
\begin{aligned}
{\m H}=-&\sum_{\langle ij \rangle}J_{1}\Sb_{i} \cdot \Sb_{j}
-\sum_{\langle  \langle ij \rangle \rangle}J_{2}\Sb_{i} \cdot \Sb_{j} \\ 
&+D\sum_{\langle  \langle ij \rangle \rangle} \nu_{ij} [\Sb_{i} \times \Sb_{j}]_{z}
-B\sum_{i}S_{i}^{z}~.
\end{aligned}
\end{equation}
In this Hamiltonian, the first two terms describe the ferromagnetic Heisenberg coupling for the nearest (we take $J_{1} > 0$) and the next nearest neighbor (we take $J_{2} >0$). The third term is Dzyaloshinskii–Moriya interaction (DMI) between the next nearest neighbor sites (the nearest neighbor DMI coupling is zero due to inversion symmetry) and the last term is the Zeeman coupling. The sign convention of the DMI ($\nu_{ij}$) is shown in Fig.~\ref{fig:Fig3}(a). The ferromagnetic ordering vector is perpendicular to the plane of the honeycomb lattice (along the $c$ axis). The spins on the $A$ and $B$ sublattice are ${\bm S}_{A}={\bm S}_{B}= S \hat{z}$ in the ground state. 

To investigate the Hamiltonian's energy spectrum and transport properties in Eq.~\eqref{Hamiltonian_model}, we employ the linear spin-wave theory, in which we map the spin operators to bosons through Holstein-Primakoff (HP) transformation. In the rest of the paper, we neglect the magnon-magnon interaction. The HP transformation is given by,  
\be 
\hat{S}^{z}_{i A} = S -  \hat{a}^{\dagger}_{i}\hat{a}_{i}, \ \   \hat{S}^{+}_{iA}=\sqrt{2S} \hat{a}_{i}, \ \  \hat{S}^{-}_{iA}=\sqrt{2S}\hat{a}^{\dagger}_{i}~, \ee 
and 
\be 
\hat{S}^{z}_{iB}=S- \hat{b}^{\dagger}_{i}\hat{b}_{i}, \ \   \hat{S}^{+}_{iB}=\sqrt{2S}\hat{b}_{i},\ \ \hat{S}^{-}_{iB}=\sqrt{2S}\hat{b}^{\dagger}_{i}~. 
\ee
Here, $\hat{a}, \hat{b}$ ($\hat{a}^{\dagger}, \hat{b}^{\dagger}$) are the magnon annihilation (creation) operator for sublattice $A$ and $B$, respectively. In addition, the spin ladder operators satisfy $S^{\pm}= S^{x} \pm i S^{y}$ for both sublattices. Further, the Fourier transform for the magnon operator in the momentum space is given by,
\begin{equation}
\begin{bmatrix}
\hat{a}_{i} \\
\hat{b}_{i}
\end{bmatrix}=\dfrac{1}{\sqrt{N}}\sum_{k} e^{i{ \bm k}\cdot{\bm r}_{i}}
\begin{bmatrix}
\hat{a}_{k} \\
\hat{b}_{k}
\end{bmatrix},
\end{equation}
where $N$ is the total number of unit cells in the lattice. Within the linear spin-wave approximation, the Hamiltonian in Eq.~\eqref{Hamiltonian_model} can be expressed in terms of the bosons in the momentum space, ${\cal H} = \sum_{\kb\in \text{1st~BZ}}{\m H}_0 (\kb)$, where, 
\be\label{eq:ham_in_k_space}  
{\m H}_0 (\kb) = \Psi^{\dagger}(\kb)H(\kb)\Psi(\kb)~. 
\ee  
{Here, $\Psi^\dagger = [{\hat a }^{\dagger} \ \ {\hat b}^{\dagger}]$}, and we have defined,
\be\label{eq:ham_tight_binding}  
H(\kb)=h_{0}(\kb)\sigma_{0}+h_{x}(\kb)\sigma_{x}+h_{y}(\kb)\sigma_{y}+h_{z}(\kb)\sigma_{z}~,
\ee  
with $\sigma_{x,y,z}$ are the Pauli matrices and $\sigma_0$ is the identy matrix of order two. In Eq.~\eqref{eq:ham_tight_binding}, the different coefficients of the $\sigma$ matrices are
\bea 
 h_{0}(\kb)& = &(3J_{1}+6J_{2})S+B \nn \\
 & & -2S\sqrt{J_{2}^2+D^2}~\cos{\phi}~\sum_{\mu} \cos(\kb\cdot \boldsymbol{a}_{\mu}), \nn  \\
 h_{x}(\kb) &=&-J_{1}S\sum_{\mu} \cos(\kb \cdot \boldsymbol{\delta}_{ \mu}), \nn \\ 
 h_{y}(\kb) &=&-J_{1}S\sum_{\mu} \sin(\kb\cdot \boldsymbol{\delta}_{\mu}),\nn\\ 
 h_{z}(\kb) &=&2S \sqrt{D^2+J_{2}^2}~\sin\phi~\sum_{\mu} \sin(\kb\cdot \boldsymbol{a}_{\mu})~.
\eea  
with $\tan \phi=D/J_{2}$. Here, $\boldsymbol{\delta}_\mu$ and $\boldsymbol{a}_\mu$, are the nearest neighbor position vectors and the lattice vector, respectively, as shown in Fig.~\ref{fig:Fig3}. Now we calculate the energy eigenvalues of this Hamiltonian and it is given by $\e_{\alpha} (\kb) = h_0(\kb) + \alpha \gamma(\kb)$ with $\gamma(\kb) = \sqrt{h_{x}^2(\kb)+h_{y}^2(\kb)+h_{z}^2(\kb)}$, and $\alpha = \pm 1$ being the two magnon branches. The energy eigenfunction  corresponding to each eigenvalue is given by,
\be
 \ket{\Psi_{\alpha}(\kb)} =\dfrac{1}{\sqrt{2}}\left( \lambda^\alpha_{+} \ \ \ \  -\alpha e^{-i\beta(\kb)} \lambda^\alpha_{-}\right)^{\rm T},
\ee 
where $ \lambda^\alpha_{\pm} = \sqrt{1  \pm \alpha {h_{z}(\kb)/\gamma(\kb)}}$, T denotes the transpose of the matrix, and  we have defined $\beta(\kb)=\arctan[h_{y}(\kb)/h_{x}(\kb)]$. Note that the eigenfunctions do not depend on $h_0 (\kb)$ at all. So, any changes in $h_0 (\kb)$ will modify only the energy eigenvalues, while the Berry curvature and other band geometric quantities remain unaltered. 

Without the DMI ($D=0$), the magnon Hamiltonian preserves both $Tc_{x}$ ($T$ being time-reversal symmetry and $c_{x}$ is a rotation of 180$^{\circ}$ around the $x$-axis in the spin space) and inversion symmetry. This makes the Berry curvature identically zero over the whole Brillouin zone with gapless Dirac points~\cite{PhysRevB.94.075401} at two nonequivalent $K$ and $K'$ points. The gapless Dirac points are also known to be robust against higher-order magnon-magnon coupling~\cite{PhysRevB.94.075401}. Including the DMI interaction breaks the $Tc_{x}$ symmetry and changes the band topology by opening a finite gap at Dirac points. This leads to a finite Berry curvature, which in turn gives rise to the linear magnon Hall effect~\cite{owerre2016topological}. The underlying physics in the presence of the DMI is similar to Haldane's quantum anomalous spin Hall model~\cite{PhysRevLett.61.2015, PhysRevLett.117.227201,owerre2016topological,cai2021topological}. Even in the absence of the DMI interaction, different quantum metric components remain finite, making this model a suitable platform for studying intrinsic nonlinear magnon transport. 

However, the presence of inversion symmetry forces all the second-order nonlinear responses to vanish, including the intrinsic contribution. This is remedied by the presence of an inversion symmetry-breaking term in the Hamiltonian which breaks the valley symmetry between $K$ and $K'$. For example, it can induce a tilt in the Dirac dispersion~\cite{PhysRevB.100.165422} by adding a term like $B_s (k_x + k_y) \sigma_0$ in Eq.~\eqref{eq:ham_tight_binding}. A similar tilt can also arise in the system, from the combined impact of the external electric field and the spin-orbit coupling via the Aharonov-Casher (AC) phase~\cite{ACphase1,ACphase3}. See Appendix.~\ref{symmetry_and_results} for details  {and a numerical estimate}. With both the inversion and time reversal symmetry broken, all the thermal current contributions become finite. 
The more interesting case is that of vanishing DMI, where the $T c_x$ symmetry gets restored, and the intrinsic nonlinear magnon thermal Hall effect is the only finite Hall response.
 We present the momentum-resolved energy dispersion of the lowest energy magnon band in Fig.~\ref{fig:density_plots}~(b) along with the Berry curvature in Fig.~\ref{fig:density_plots}~(c). The quantum metric tensor has three distinct components that are shown in Fig.~\ref{fig:density_plots}~(d)-(f). We can clearly see that the band geometric quantities are highly concentrated near the band edges at the $K,\ K'$ points having coordinates $(\pm k_0,0)$, with $k_0 = 4\pi/3\sqrt{3}$. It is instructive to study the low-energy Hamiltonian around these band edges. With the help of the Taylor series expansion, we obtain the low-energy Hamiltonian for Eq.~\eqref{Hamiltonian_model} near $K, K'$ points,
\begin{equation}\label{eq:low_energy_ham}
H_{\zeta}=h_{0}\sigma_{0}+v_{m}(\zeta k_{x}\sigma_{x}+k_{y}\sigma_{y})+\zeta \Delta \sigma_{z}~,
\end{equation}
where, $\zeta=\pm 1$ for $K$ and $K'$ points, and $k_x/k_y$ are measured with respect to the corresponding $K$ and $K'$ points. The low-energy magnon Hamiltonian is analogous to the massive tilted Dirac Hamiltonian  ~\cite{du_NatCom2019_disorder} in two-dimensional hexagonal fermionic systems with broken inversion symmetry. However, there is one crucial difference. For the fermionic system, the mass term that opens up a gap near the $K, K'$ points is independent of the valley index. This results in the total Chern number being zero for each band in fermionic systems with time-reversal symmetry. In contrast, for our magnon Hamiltonian, the DMI-induced gap term has an opposite sign near $K-K'$, so the gap opening is a topological transition with the magnon bands having a finite Chern number. To have a relatively simplified expression, we set the next-to-next nearest neighbor interaction ($J_2$) to be vanishingly small. Consequently, the parameters of the Hamiltonian have the following expression: $h_0 = 3J_1S + B + B_s( k_x + k_y)$, $v_m = 3J_1 S/2$, and $\Delta = - 3\sqrt{3} S D$ arises from the DM interaction and opens a gap in the magnonic spectrum. The energy eigenvalues are given by $\e_{\alpha} = h_0 + \alpha \e_0$, with $\e_0 = \sqrt{(k_x^2 + k_y^2) v^2_m + \Delta^2}$, independent of the valley index. 
The corresponding eigenfunctions are, 
%
%
\be  
\ket{\psi_{\alpha} (\kb) }_{\zeta}= \frac{1}{\sqrt{2}} \left(\alpha \sqrt{1+\alpha \frac{\zeta \Delta}{\e_0} } e^{-i \theta_\zeta} \ \ \ \ \sqrt{1-\alpha \frac{\zeta \Delta}{\e_0} }\right)^{\rm T}~, 
\ee%
where we have defined $\theta_\zeta = \arctan(k_y /\zeta k_x)$. Using these  simplifications, we calculate different band geometric quantities. These are given by, 
%
\be\label{eq:lowenergy_qgc} 
\bal   
 & {\Omega}^{xy}_{\alpha} = - \frac{\alpha \Delta v_m^2}{2\e^3_0}~, \\
 & {\m G}^{xx}_{\alpha,-\alpha} = \frac{v_m^2(\Delta^2 + k_y^2 v_m^2)}{4\e_0^4}~,\\
 & {\m G}^{xy}_{\alpha,-\alpha} = -\frac{k_x k_y v_m^4}{4\e_0^4}~, \\
 & {\m G}^{yy}_{\alpha,-\alpha} = \frac{v_m^2(\Delta^2 + k_x^2 v_m^2)}{4\e_0^4}~.
\eal
\ee 
%
Note that the magnon band geometric quantities are independent of the valley index, though the Berry curvature depends on the band index. 
We highlight a few things here. Firstly, the Berry curvature is zero in the absence of the DMI, as it is directly proportional to the $\Delta$. Secondly, the tilting term ($B_s(k_x+k_y)\sigma_0$) introduced in the Hamiltonian with an identity matrix does not influence any of the band geometric quantities but invalidates the relation $\epsilon_{\alpha}(\bold{k})=\epsilon_{\alpha}(-\bold{k})$, as expected. In the Appendix.~\ref{symmetry_and_results} we discuss the possible microscopic mechanism responsible for such a term in a spin model. Lastly, in the absence of DMI and external magnetic field, the lowest energy magnon dispersion reduces to the gapless magnon excitation when $\kb \to 0$. This gapless Goldstone mode is the consequence of the spontaneously broken symmetry in the ferromagnetic ground state.
Furthermore, the calculated expression of the Berry curvature is consistent with Refs.~[\onlinecite{du_PRL2018, 
 carmine_AQT2021, harsh_arx2022_quantum}], and the distinct components of the quantum metrics are in complete agreement with Refs.~[\onlinecite{huiying_PRL2021_intrinsic,harsh_arx2022_quantum}].  


\section{NONLINEAR MAGNON THERMAL HALL
EFFECT in Hexagonal ferromagnets}\label{results and discussion}
\begin{figure}[t!]
\centering
\includegraphics[width=\linewidth]{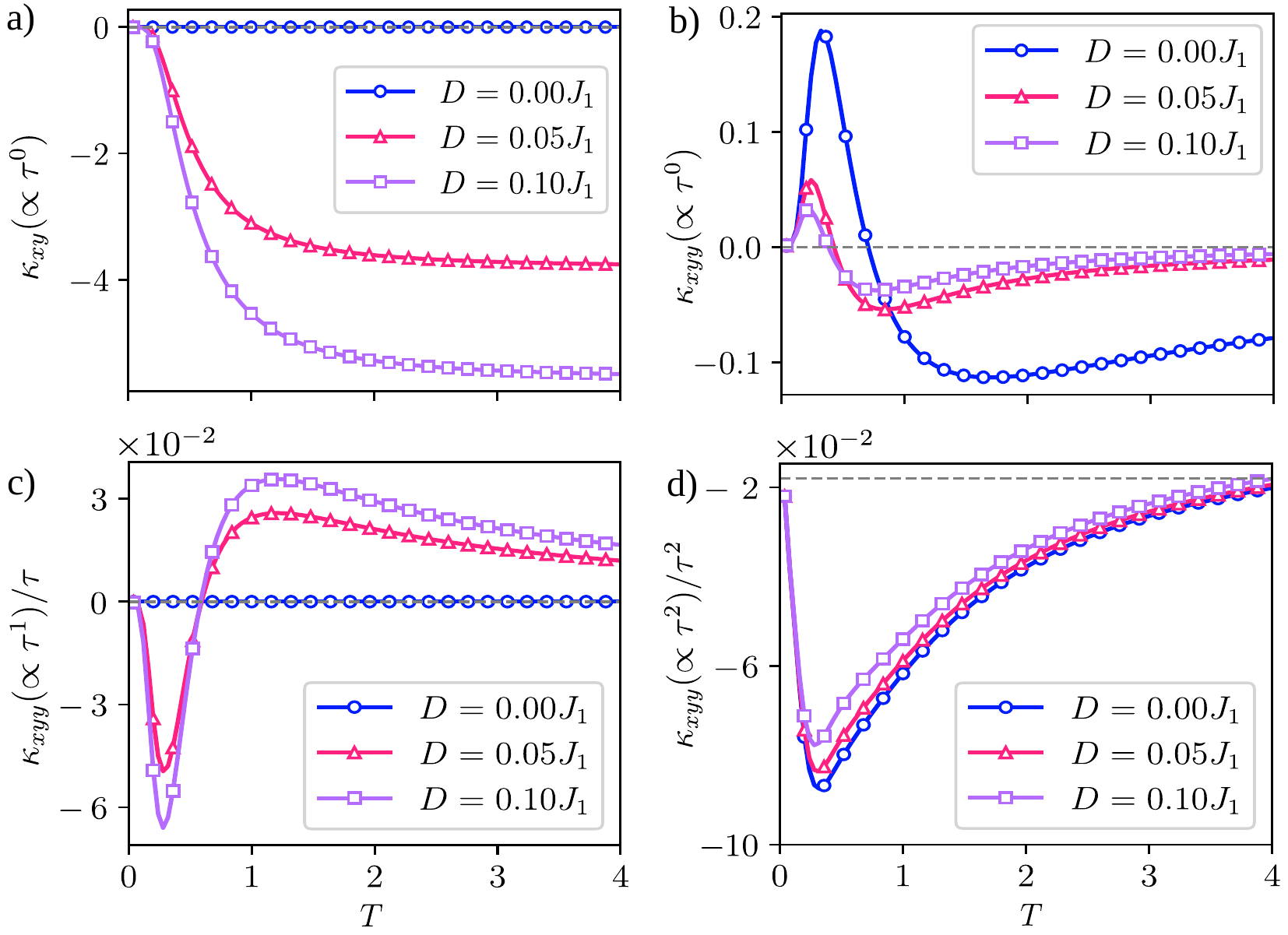}
\caption{Thermal Hall conductivity for magnons. a) the linear, b) the nonlinear intrinsic, c) the nonlinear extrinsic anomalous, and d) the nonlinear Drude contribution to the thermal Hall conductivity for three values of the Dzyaloshinskii-Moriya interaction (DMI). 
The Berry curvature dependent linear Hall and the extrinsic nonlinear anomalous component of the thermal current are induced by the presence of a finite DMI, and vanish as DMI vanishes. In contrast, the intrinsic nonlinear contribution increases with decreasing gap or DMI and is finite even in the vanishing DMI limit. The Drude contribution does not depend on any band geometric quantity. In the limit of vanishing DMI, the intrinsic Hall contribution dominates over all other contributions. 
We have chosen the same parameters as in Fig.~\ref{fig:Fig3} except $B_{s}= 0.01 J_{1}$. In these plots, we have used the natural units and set $\hbar =1 $ and $k_B = 1$. The grey dashed line marks the $0$ and serves as a guide to the eye.}
    \label{fig:magnon_current}
\end{figure}

We present the variation of the linear and nonlinear magnon transport coefficients with temperature in Fig.~\ref{fig:magnon_current} for different values of DMI strength. First, let us focus on the linear thermal Hall conductivity  depicted in Fig.~\ref{fig:magnon_current}(a). As the linear thermal Hall conductivity is directly proportional to the Berry curvature, its value increases with increasing DMI coupling. For zero DMI, the Berry curvature becomes zero and so does the Hall conductivity. The linear thermal Hall current doesn't show any sign change as a function of temperature or the DMI strength. The results are consistent with the study in Ref.~[\onlinecite{owerre2016topological}]. 

In Fig.~\ref{fig:magnon_current}(b), we  show the temperature dependence of the intrinsic nonlinear thermal Hall conductivity for several values of DMI. Recall that the intrinsic nonlinear contribution depends solely on the different components of the quantum metric. From Eq.~(\ref{eq:lowenergy_qgc}), we can clearly see that all the components of the quantum metric  are maximum when the DMI is zero (${\cal G}\propto \Delta^{-2}$, at the band edge). This is also reflected in the plot of the nonlinear intrinsic current. 

In Fig.~\ref{fig:magnon_current}(c), we plot the Berry curvature dipole-induced extrinsic nonlinear Hall conductivity which varies linearly with the magnon scattering time ($\tau$). Note that the extrinsic nonlinear anomalous Hall conductivity vanishes as the DMI strength goes to zero. In Fig.~\ref{fig:magnon_current}(d), we present the Drude contribution to the nonlinear Hall effect. 
This contribution is independent of any band geometric quantities like the Berry curvature or the quantum metric, it entirely depends on the band dispersion and the magnon distribution function. This can also be seen from the fact that the curves for different DMI values are very similar in nature. 

 {To understand the behaviour of these responses at high temperatures, we note the thermal driving force (akin to the electric field for fermions) is $\propto \nabla T/T$. Thus, a constant temperature gradient is less effective at higher temperatures, with the thermal fluctuations nullifying the impact of the temperature gradient. Consequently, larger thermal fluctuations suppress the directed flow of magnons under a fixed temperature gradient as $T$ increases. This suppression is more dominant in the nonlinear thermal responses as compared to linear responses.}
\begin{table*}[t]
\centering 
\begin{tabular}{ cccc } 
\hline
\hline
\rule{0pt}{3ex}
 DMI strength (in unit of $J_{1}$) & $\tau$ dependence  & Linear response (W/m)& Nonlinear response (W/m) \\ 
\hline
\hline
\rule{0pt}{3ex}
\multirow{3}{4.1cm}{\centering $D = 0.0$} & $\tau^{0}$ & $0.0$ & $4.3 \times 10^{-16}$\\ 
& $\tau^{1}$ & 0.0 & 0.0 \\
&  $\tau^{2}$ & - & $7.8 \times 10^{-16}$\\  
\hline
\rule{0pt}{3ex}
\multirow{3}{4.1cm}{\centering $D=0.1$}  & $\tau^{0}$ & $2.75 \times 10^{-9}$ & $3.6 \times 10^{-17}$\\ 
    & $\tau^{1}$ & 0.0 & $8.8 \times 10^{-17}$\\ 
           & $\tau^{2}$ & - & $1.6 \times 10^{-16}$\\ 
\hline
\hline
\end{tabular}
\caption{ Demonstration of the magnitude of linear and nonlinear magnon thermal Hall current at a temperature of 20 K for different values of the DMI.
In calculating these numerical values, we used the same parameters as in Fig.~\ref{fig:density_plots} except $B_s = 0.01$ $J_1$, and we assume the magnon lifetime to be of the order of $\tau \sim 1$ pico-seconds. 
\label{magnetude}}
\end{table*}

 

        


\section{Discussions}
\label{Discussions}

The potential material candidates which host magnons in a hexagonal ferromagnetic lattice are chromium trihalides such as $\ce{CrBr_3}$~\cite{crbr3}, $\ce{CrI_3}$~\cite{cri3}, and $\ce{CrCl_3}$~\cite{crcl3}. These compounds are promising materials for probing band geometry-induced nonlinear thermal transport in magnons. Recent measurements on $\ce{CrCl_3}$~\cite{crcl3exp} suggest the presence of gapless Dirac-like magnon dispersion. However, the magnon dispersion of $\ce{CrBr_3}$~\cite{crbr3exp} and $\ce{CrI_3}$~\cite{cri3exp} indicates the presence of a gap  at $K$ point. This suggests that the effective DMI interaction for $\ce{CrCl_3}$ is zero. As a consequence, $\ce{CrCl_3}$ is a potentially good platform to study the nonlinear intrinsic magnon thermal Hall response highlighted in Fig.~\ref{fig:magnon_current}(b). 

To compare the relative magnitude of different linear and nonlinear contributions to the thermal Hall current, we present an order of magnitude estimation of different contributions in Table~\ref{magnetude}. For the analysis, we assume the magnon lifetime to be $\tau \sim 1$ pico-seconds~\cite{lifetimecri3,lifetime1} which is the typical value for chromium trihalide ferromagnets. The  temperature gradient used for a magnon transport experiment  in Ref.~[\onlinecite{PhysRevB.96.134425}] was of the order of $\nabla T \sim 10^{-6}$ K/nm. In comparison, the magnitude of the linear magnon thermal Hall current observed in $\ce{Lu_2V_2O_7}$ (ferromagnetic insulator) bulk sample at $T = 20$ K was $4 \times 10^{-4}$ W/K.m~\cite{magnetude1}.  Assuming the thickness of the monolayer sample to be 0.5nm,  the thermal Hall conductivity for single monolayer is $2 \times 10^{-13}$ W/K~\cite{magnetude2,magnetude3}. Assuming the same temperature gradient for our case, we find the corresponding thermal Hall current to be of the order $2 \times 10^{-10}$ W/m.

 {
An interesting issue from the experimental perspective is to separate the intrinsic contribution ($\propto \tau^0$) and the Drude contribution ($\propto \tau^2$) in the measured nonlinear thermal Hall effect. A potential solution is to plot the measured total nonlinear transverse thermal conductivity as a function of linear longitudinal conductivity ($\kappa_{L} \propto \tau$) by varying some system parameters such as temperature or strain. The intrinsic part of the nonlinear conductivity will scale as $(\kappa_{L})^0$, and the extrinsic Drude part will scale as $(\kappa_{L})^2$. A similar approach helped in identifying the intrinsic and extrinsic electrical transport contributions to the nonlinear conductivity (for fermions) in Ref.~[\onlinecite{wang_quantum_2023}], and in Ref.~[\onlinecite{gao_quantum_2023}]. }

At finite temperature, the magnon-magnon interaction becomes important and it can renormalize the magnon dispersion along with the corresponding band geometric quantities and the magnon relaxation time \cite{magnonlifetimep}. It was shown in Ref.~[\onlinecite{magnonlifetime1}] that for ferromagnetic monolayer $\ce{Cr_2Ge_2Te_6}$, the relaxation time increases with temperature ($\tau \propto T^2$) due to the magnon scattering 
while it decreases with 
the external magnetic field. 
In our study, for simplicity, we have assumed the relaxation time to be a constant and ignored the effects of magnon-magnon interaction. The investigation of the impact of magnon-magnon interaction on different nonlinear transport coefficients can be an interesting study for the future. 

 {At larger lengthscales, the dipolar interaction can also play an important role and modify the dispersion relation, spin-orbit coupling, magnon-magnon coupling, etc. However, in this manuscript, our focus is on magnetic systems where the exchange interactions dominate over the dipolar interactions, allowing us to disregard the latter~\cite{sorino}. This is a reasonable regime for ferromagnetic materials like CrCl$_3$.}

 {A critical aspect of our study is that it is valid in the weak disorder limit, where the quasiparticle approximation with extended wavefunction is still reasonable, with ${\bm k}$ being a good quantum number. In the limit of large disorder density and disorder strength, new physics may emerge (like localization, etc.), where a completely different framework and analysis will be needed.}

 {Several earlier works have established that $\nabla T$ can be treated as a physical force in the linear response regime for fermionic systems. Luttinger \cite{luttinger_PR1964_theory} first treated $\nabla T$ as a scalar potential, and  Tatara \cite{tatara_PRL2015_thermal} treated $\nabla T$ as a vector potential to capture its impact on the thermal transport properties of fermions. More recently, Ref.~\cite{nagaosa_PRB2020_quantum} showed that using the quantum kinetic theory and treating $\nabla T$ as a vector potential reproduces all the known thermal Hall and other responses induced in the linear response regime for fermions. Motivated by this, we showed in Ref.~\cite{harsh_arx2022_quantum} that the treatment of $\nabla T$ as a vector potential can be extended to the nonlinear response regime for fermions and this framework reproduces all the known results from the Boltzmann transport theory which treats $\nabla T$ as a statistical force. These works establish that treating $\nabla T$ as a vector potential captures the linear and nonlinear thermal response of fermions. Our work demonstrates this idea for bosons. Since the single-particle quantum mechanics 
of bosons and fermions are identical, they have very similar band geometric quantities. Thus, within the quantum kinetic theory, the linear and nonlinear heat transport by bosons and fermions have very similar expressions, with their difference being captured only by their equilibrium statistics \cite{harsh_arx2022_quantum}. This should work as long as we can treat the fermions and bosons at a single particle level. This may not work for strongly interacting systems where a multi-particle framework will be needed, and the differing statistics of fermions and bosons may play an important role.}



\section{Conclusion}
\label{Conclusion}
In conclusion, we have developed a quantum kinetic theory framework for systematically studying thermal transport in bosonic systems up to second-order in the applied temperature gradient. Our theoretical framework provides an alternative to the 
Boltzmann transport theory. It includes all interband coherence effects and works without the need 
for boundary confining potentials. 

We predict the existence of an intrinsic nonlinear boson thermal current, which arises from the quantum metric and is independent of the scattering time. In contrast, the extrinsic contributions, such as the anomalous and Drude parts of the nonlinear thermal current, exhibit linear and quadratic dependencies on the scattering time, respectively. Considering topological magnons in a two-dimensional ferromagnetic honeycomb lattice as a concrete example, we show that the intrinsic nonlinear magnon thermal Hall current dominates over other responses in the absence of DMI. The absence of DMI ensures that the Berry curvature and the linear Hall response vanishes along with the nonlinear anomalous thermal Hall response.  

Our study highlights the significance of band geometry-induced nonlinear thermal transport. It opens up new avenues for experimental probe of intrinsic nonlinear thermal Hall response of bosons.  {A potentially interesting problem is to develop framework for probing the signature of real-space topology in magnon transport \cite{jin2023nonlinear}, similar to topological Hall effect for fermions \cite{Shao_THE_Nature19,Verma_sciadv.abq2765}.}
Our findings have potential implications for quantum magnonics, and our calculations can be easily extended to explore other bosonic systems. 

\section{ACKNOWLEDGEMENTS}
R.~M. is grateful for useful communication with Ran Cheng (University of California, Riverside), Robin R. Neumann (University of Halle), and Saikat Bannerjee (Los Alamos). R.M. acknowledges the CSIR (Govt. of
India) for financial support. R.M is also supported by a Fulbright Program grant under the Fulbright-Nehru Doctoral Research Fellowship, sponsored by the Bureau of Educational and Cultural Affairs of the United States Department of State and administered by the Institute of International Education and the United States-India Educational Foundation. H. V. thanks the MHRD, India for funding through the Prime Minister’s Research Fellowship (PMRF).  We  acknowledge the Department of Science and Technology of the Government of India, for financial support via Project No. DST/NM/TUE/QM-6/2019(G)-IIT Kanpur.

\onecolumngrid
\appendix
\section{Calculation of the second-order density matrix}\label{app:cal_of_so_dm}
To calculate the second-order density matrix, we use recursive Eq.~\eqref{eq:n-th_dm}, and write the second-order density matrix in the following form,
\be\label{eq:so_dm_main}  
\rho^{(2)}_{np} = -i\hbar g^{np}_2 \left[ D_T (\rho^{(1)})\right]_{np}~.
\ee  
For ease of calculation, we segregated $\rho^{(2)}$ into the four parts as $\rho^{\rm{dd}},~\rho^{\rm{do}},~\rho^{\rm{od}},~{\rm{and}}~\rho^{\rm{oo}}$. Here, $\rho^{\rm{dd}}$ and $\rho^{\rm{do}}$ are diagonal matrices in the band index, while $\rho^{\rm{od}}~{\rm{and}}~\rho^{\rm{oo}}$ are off-diagonal matrices. So, to calculate the diagonal part of $\rho^{(2)}$, we calculated the diagonal elements of the thermal driving term to be  
\be
[D_T(\rho^{(1)})]_{nn} = -\frac{1}{2\hbar} E^b_T \left[ 2 \e_n \partial_{b} \rho^{\rm{d}}_{nn} + i \sum_{p \neq n } (\e_n + \e_p) ( \rho^{\rm{o}}_{np} {\m{R}}^{b}_{pn} - {\m{R}}^{b}_{np} \rho^{\rm{o}}_{pn}) \right].
\ee
On the basis of the classification discussed in Sec.~\ref{sec:so_dm}, the `dd' part of $\rho^{(2)}$ arising from the first term of the above equation is given as 
\be
\rho^{\rm{dd}}_{nn} = - \frac{\tau}{ 2\hbar }\e_n \partial_b(\rho^{\rm{d}}_n)E^b_T =  \frac{\tau^2}{ 2\hbar }\e_n \left[ v^n_b \partial_c f^B_n + \frac{\e_n}{\hbar} \partial_b\partial_c f^B_n\right] E^b_T E^c_T~,
\ee
and the remaining part of $\rho^{(2)}$, which is diagonal in band index and originates from $\rho^{\rm o}$ is given by 
\be 
 \rho^{\rm{do}}_{nn}  = \frac{i\tau}{4\hbar}\sum_{p} (\e_n + \e_p) \left(g^{np}_1\m{R}_{np}^{c}\m{R}_{pn}^{b} + g^{pn}_1\m{R}_{np}^{b}\m{R}_{pn}^{c}\right)\xi_{np} E^b_T E^c_T~.
\ee  
Following a similar approach, we now calculate the remaining two parts of $\rho^{(2)}$ stemming from the off-diagonal elements of the thermal driving term. After performing a careful calculation, we obtain the off-diagonal ($n \ne p$) elements of the thermal driving term as,
\be
[D_T(\rho^{(1)})]_{np}= -\frac{1}{2\hbar} E^b_T\bigg[ (\e_n + \e_p)\partial_{b}\rho^{\rm{o}}_{np} + i \sum_{q} \bigg( (\e_n + \e_q)\rho^{(1)}_{nq}\m{R}^{b}_{qp} - (\e_q + \e_p)\m{R}^{b}_{nq}\rho^{(1)}_{qp} \bigg) \bigg].
\ee
The second term inside the square bracket of the above equation containing summation over $q$ can be simplified by considering the following three cases:--- $q = n \neq p $, $q = p \neq n $, and $q \neq n \neq p$. 
Thus, we simplify the above expression in the form of
\be 
[D_T(\rho^{(1)})]_{np} = -\frac{1}{2\hbar} E^b_T\bigg[ 2 i (\e_n \rho^{\rm{d}}_{nn}  -  \e_p \rho^{\rm{d}}_{pp})\m{R}^b_{np} + (\e_n + \e_p)\m{D}^b_{np} \rho^{\rm{o}}_{np} + \sum_q^{q \neq n \neq p } i \left( (\e_n + \e_q)\rho^{\rm{o}}_{nq}\m{R}^{b}_{qp} - (\e_q + \e_p)\m{R}^{b}_{nq}\rho^{\rm{o}}_{qp} \right) \bigg]~,
\ee
where we have defined $\m{D}^b_{np} = \partial_b - i ( \m{R}^b_{nn} - \m{R}^b_{pp})$. 
Using this expression of the thermal driving term in Eq.~\eqref{eq:so_dm_main}, we calculate the off-diagonal elements of $\rho^{(2)}$ originating from $\rho^{\rm d}$ as 
%
\be
\rho^{\rm{od}}_{np} = - E^b_T g_{np} \m{R}^b_{np} (\e_n \rho^{\rm{d}}_{nn} - \e_p \rho^{\rm{d}}_{pp}) =  \frac{\tau}{\hbar}g_2^{np} \m{R}_{np}^{b}\left(\e_n ^2 \partial_{c}f^B_{n}-\e_p^2 \partial_{c}f^B_{p} \right)  E^b_T E^c_T~.
\ee  
Finally, we left out with $\rho^{\rm oo}_{np}$, which corresponds to the off-diagonal part of $\rho^{(2)}$ stemming from $\rho^{\rm o}$. Explicit calculation of this part results in the following form,
\bea
 \rho^{\rm{oo}}_{np}  &=& -\frac{i}{2}g^{np}_2 (\e_n + \e_p ) \m{D}^{b}_{np}\left(g_{np}\m{R}_{np}^{c}\xi_{np}\right) E^b_T E^c_T  \nn \\ 
&& \  +  \frac{1}{ 2}g^{np}_2 \sum_{q\neq n \neq p} \left[ g^{nq}_1 \m{R}^{c}_{nq}\m{R}^{b}_{qp}(\e_n +\e_q )\xi_{nq} - g^{qp}_1 \m{R}^{b}_{nq}\m{R}^{c}_{qp}(\e_q +  \e_p )\xi_{qp}\right]E^b_T E^c_T~,
\eea
Hence, we completed the calculation of the second-order density matrix in response to the applied temperature gradient. 
\section{Linear thermal current}\label{app:linear_th_cur}
In this section, we present the calculation of the linear thermal current within the quantum kinetic theory framework. Starting from Eq.~\eqref{eq:th_cur_def}, we show that the first term on the right-hand side of this equation can be written as 
\be\label{eq:fo_th_cur_cal}   
\bal   
{\rm{Tr}}\bigg[\frac{1}{2}\{\m{H}_0,\bm{v} \}\rho^{(1)}\bigg] &= \frac{1}{2} \sum_{p,\kb} \langle u^p_\kb \vert {\m H}_0 {\bm v} \rho^{(1)} + {\bm v} {\m H}_0 \rho^{(1)]} \vert u^p_\kb \rangle \\
&= \sum_{n,p,\kb} (\e_n 
 + \e_p) {\bm v}_{pn} \rho^{(1)}_{np}~, \\
 &= \frac{1}{2} \sum_{n,p,\kb} (\e_n + \e_p) ({\bm v}_n \delta_{pn} + i\omega_{pn} {\bm {\m R}}_{pn}) (\rho^{\rm d}_{nn} \delta_{np} + \rho^{\rm o}_{np})~, \\ 
 &= \frac{1}{2} \sum_{n,p,\kb} (\e_n + \e_p) ( {\bm v}_n \rho^{\rm d}_{nn} \delta_{pn} + i \omega_{pn}{\bm {\m R}}_{pn}\rho^{\rm o}_{np} )~. 
\eal   
\ee   
Here, we use the elements of the first-order density matrix in the form of $\rho^{(1)}_{np} = \rho^{\rm d}_{nn} \delta_{np} + \rho^{\rm o}_{np}$. In the last line of the above equation, we consider that the $\omega_{pn}$ and $\rho^{\rm o}_{np}$ are zero when $n = p$. 
Here, we see that the first term of Eq.~\eqref{eq:th_cur_def} gives two contributions to thermal current---one originating from $\rho^{\rm d}$ and the other from $\rho^{\rm o}$. So, the linear thermal current component along the $a$ axis arising from the $\rho^{\rm d}$ will be given as
\be   
J^{\rm d}_{a} = \sum_{n,\kb} \e_n v^a_n \rho^{\rm d}_{nn} = -\frac{\tau}{\hbar} \sum_{n,\kb} \e_n^2 v^a_n \partial_c f_n^B E^c_T~.
\ee   
This thermal current component depends entirely on the energy dispersion and the distribution function. Therefore, it is also called the ``linear Drude thermal current.''
Similarly, another thermal current contribution arising from $\rho^{\rm o}$ is given as 
 \be 
 \bal  
J^{\rm o}_a &= \frac{i}{2\hbar} \sum_{n,p,\kb} (\e_n + \e_p) \e_{pn} {\m R}^a_{pn} \rho^{\rm o}_{np}~, \\
&=  -\frac{i}{2\hbar} \sum_{n,p,\kb} (\e_n + \e_p) \e_{pn} g_1^{np} {\m R}^a_{pn} {\m R}^c_{np} (\e_n f_n^B- \e_p f_p^B) E^c_T~.
\eal 
 \ee  
Using the dilute impurity limit, which we will discuss in Sec.~[\ref{app_C}], we can write $g^{np}_1 \approx 1/\e_{np}$. With this substitution, the above equation can be simplified to 
\be\label{B4}  
\bal   
J^{\rm o}_a &= \frac{i}{2\hbar} \sum_{n,p,\kb} (\e_n + \e_p) {\m R}^a_{pn} {\m R}^c_{np} (\e_n f_n^B- \e_p f_p^B) E^c_T~, \\
&= -\frac{1}{2\hbar} \sum_{n,p,\kb} (\e_n + \e_p) \e_n f_n^B \Omega^{ac}_{np} E^c_T~. 
\eal  
\ee   
In writing the last line of the above equation, we have used some manipulation:--- firstly, we decomposed the first line of the above equation into two terms through $(\e_n f_n^B- \e_p f_p^B)$. Secondly, we shuffled the dummy indices $n,p$ in the latter part of the decomposition. Finally, we grouped the terms and used the definition of the band-resolved Berry curvature. Henceforth, we will use this manipulation to write the thermal current in terms of the band-resolved Berry curvature and the quantum metric. Due to the presence of the Berry-curvature, this component of the linear thermal current is always perpendicular to the applied temperature gradient. Hence, it gives rise to the contribution to the linear thermal Hall response. 

The second term on the right-hand side of eq.~\eqref{eq:th_cur_def} gives the particle magnetic moment-induced contribution to the linear thermal current, which can be written as  
\be\label{eq:fo_mag_cur}   
{\rm{Tr}}\left[(\bm{E}_T \times \bm{m}_N) \m{H}_0 \rho_0 \right]_a = \sum_{n,\kb}\epsilon_{acb} E^c_T m^{b}_{N,n} \e_n f^B_n~.
\ee  
Here, $\epsilon_{acb}$ is the antisymmetric Levi-Civita tensor of rank 3 which $(a,b,c)$ being the even cyclic permutation. Using the expression of the particle magnetic moment given in Eq.~\eqref{eq:pmm}, we can show 
\be 
\bal  
{\bm m}_{N,n} &= -\frac{i}{2\hbar} \sum_{p \ne n} (\e_n - \e_p) {\bm {\m R}_{np}} \times {\bm {\m R}_{pn}}~, \\
\implies m^a_{N,n} &=  -\frac{i}{2\hbar} \sum_{p \ne n } \sum_{b,c} \epsilon_{abc} (\e_n -\e_p) {\m R}^b_{np} {\m R}^c_{pn} ~, \\
\implies \epsilon_{abc} m^a_{N,n} &= -\frac{1}{2\hbar} \sum_{p \ne n} (\e_n - \e_p) \Omega^{bc}_{np}~.
\eal 
\ee   
Using the fact that $\epsilon_{acb} = - \epsilon_{abc} = \epsilon_{bac}$ along with the above definition of the particle magnetic moment in Eq.~\eqref{eq:fo_mag_cur}, we get the magnetic moment contribution of the linear thermal current as 
\be  
 {\rm{Tr}}\left[(\bm{E}_T \times \bm{m}_N) \m{H}_0 \rho_0 \right]_a = -\frac{1}{2\hbar} \sum_{n,p,\kb} \e_n (\e_n -\e_p) \Omega^{ac}_{np} f^B_n E^c_T~.
\ee  
This contribution also gives a Hall-like response to the linear thermal current. Further, this contribution is similar to $J^{\rm o}$. After combining these two terms, we write
\be  
J^{\rm o}_a + {\rm{Tr}}\left[(\bm{E}_T \times \bm{m}_N) \m{H}_0 \rho_0 \right]_a  = -\frac{1}{\hbar} \sum_{n,\kb} \e_n^2 \Omega^{ac}_n f^B_n E^c_T~. 
\ee  
In obtaining this result, we used a relation connecting band-resolved Berry curvature to single band Berry curvature via $\Omega^{ac}_{n} = \sum_{p \ne n} \Omega^{ac}_{np}$. 

Finally, we focus on the last term on the right-hand side of Eq.~\eqref{eq:th_cur_def}. We transform it as
\be  
\bal  
2{\rm Tr}[\bm{E}_T \times \bm{M}_{\bm \Omega}]_a & = -2{\rm Tr}[\bm{M}_{\bm \Omega} \times \bm{E}_T]_a  \\ 
&= -2 \sum_{n,\kb} \epsilon_{abc} M^b_{n,\Omega}E^c_T ~, \\ %
& =  \frac{2}{\hbar} \sum_{n,\kb} \epsilon_{abc}  \Omega^b_n(\kb) E^c_T \int^{\infty}_{\e_n} d\e (\e-\mu) f^B(\e)~, \\ 
& = -\frac{2}{\hbar} \sum_{n,\kb} \Omega^{ac}_n(\kb) E^c_T \int^{\infty}_{\e_n} d\e (\e-\mu) f^B(\e)~.
\eal  
\ee  
In the last line of the above equation, we use mathematical manipulation:--- $\epsilon_{abc} \Omega^{b}_n = \epsilon_{bca} \Omega^b_{n} \equiv \Omega^{ca}_n = -\Omega^{ac}_n$. After performing the integration along with some manipulations, we can show that
\be  
\int^{\infty}_{\e_n} d\e (\e-\mu) f^B(\e) = \frac{1}{2}(k_B T)^2 \left[ \log(1+f_n^B)]\log(\frac{1+f_n^B}{(f^B_n)^2}) - 2{\rm Li}_2(-f_n^B) \right]~.
\ee 
Here, Li$_2$ is the polylogarithmic function~\cite{lewin_book1991_structural} of order 2. Again, the contribution to the linear thermal current emanating the Berry curvature-induced heat magnetization is perpendicular to the temperature gradient. Thus, the total linear anomalous thermal Hall current is the sum of $J^{\rm o}$, ${\rm{Tr}}\left[(\bm{E}_T \times \bm{m}_N) \m{H}_0 \rho_0 \right]$  and $2{\rm Tr}[\bm{E}_T \times \bm{M}_{\bm \Omega}]$. Adding these three contributions, we have the simplified form of the linear anomalous thermal Hall current as  
\be  
\bal  
J^{\rm A}_a &= J^{\rm o}_a + {\rm{Tr}}\left[(\bm{E}_T \times \bm{m}_N) \m{H}_0 \rho_0 \right]_a + 2{\rm Tr}[\bm{E}_T \times \bm{M}_{\bm \Omega}]_a~, \\ 
&= - \frac{(k_B T)^2}{\hbar}\sum_{n,\kb} \Omega^{ac}_n \left[ \beta^2\e_n^2 f^B_n + \log(1+f_n^B)\log(\frac{1+f_n^B}{(f^B_n)^2}) - 2{\rm Li}_2(-f_n^B)  \right] E^c_T 
\eal 
\ee 
It is straightforward to show that $\beta \e_n \equiv \log(\frac{1+f^B_n}{f^B_n})$. Using this substitution, we simplify the above equation as 
\be  
J^A_a = -\frac{(k_B T)^2}{\hbar}\sum_{n,\kb} \Omega^{ac}_n c_2(f^B_n) E^c_T~,
\ee  
where $c_2(f) = (1+f)\log(\frac{1+f}{f})^2 - \log(f)^2 - 2{\rm Li}_2(-f)$. The superscript `A' in the above equation is used for linear anomalous current. This expression of the linear anomalous thermal Hall current is in complete agreement with the existing work done in Refs.~[\onlinecite{matsumoto_PRL2011_theoretical,park_NanoLett2020_thermal, neumann_PRL2022_thermal}], where it has been calculated within the semiclassical framework. 

\section{\label{app_C} Intrinsic and extrinsic nonlinear thermal currents}
From the main text, we note that the scattering time dependence of the contributions ${\bm J}^{\rm dd}$ and ${\bm J}^{\rm mag}$ are $\tau^2$ and $\tau$, respectively. However, the scattering time dependence of the remaining current is not trivial and is determined by the factor $\tau g_{np}$ for ${\bm J}^{\rm do}$ and ${\bm J}^{\rm od}$, and $g_{np}$ for ${\bm J}^{\rm oo}$. We use an identity given in Ref.~[\onlinecite{harsh_arx2022_quantum}] to extract the scattering time dependence of these factors. Following these identities, we write
\be\label{eq:identity}
\tau g^{np}_N = \frac{i\hbar}{\e^2_{np}}\left( N + \eta_{\tau,N}^{np}\right) ~~\mbox{and}~~g^{np}_N  = \frac{1}{ \e_{np}}(1 + \tilde{\eta}^{np}_{\tau,N}) .
\ee
Here, $\eta^{np}_{\tau,N}$ and $\tilde{\eta}^{np}_{\tau,N}$ are dimensionless function of $\tau\omega_{np}$ with $\omega_{np} = \e_{np}/\hbar$ being the interband transition frequency. The explicit expression of these functions is given as
\be
\bal
& \eta^{np}_{\tau,N}=  -i \tau \omega_{np} \left( \frac{ 1 -i N^3 \frac{1}{\tau^3 \omega_{np}^3} }{1 + N^2 \frac{1}{\tau^2 \omega^2_{np}}} \right)~,  \\
& \tilde\eta^{np}_{\tau,N} = N\frac{i}{\tau \omega_{np}} \left(\frac{1 + N \frac{i}{\tau \omega_{np}}}{1 + N^2 \frac{1}{\tau^2 \omega^2_{np}}}\right)~.
\eal
\ee
%
%
{In the above expressions, we emphasize that $n \ne p$. With these expressions, the form of the nonlinear thermal conductivities becomes convoluted. Therefore, to represent the nonlinear thermal conductivities in a simplified and more tractable form, we employ the dilute impurity limit (DIL), which corresponds to the scattering time ($\tau$) being greater than the inverse of the interband transition frequency $\omega_{np}$, $i.e., \ \tau \gg \frac{1}{\omega_{np} } \ {\rm or} \ \tau \omega_{np} \gg 1$.}
In DIL, we can approximate dimensionless functions $\eta_{\tau, N}^{np}$ and  ${\tilde \eta}_{\tau, N}^{np}$ straightforwardly as:-- $\eta_{\tau, N}^{np} \approx -i\tau \omega_{np}$ and ${ \tilde \eta}_{\tau, N}^{np} \approx 0 $, which in turn give $\tau g^{np}_N \approx \frac{i\hbar N}{\e^2_{np}} + \frac{ \tau}{\e_{np}}$ and $g^{np}_N \approx \frac{i}{\e_{np}}$. 
%
From the approximated form of the scattering time-dependent factor $\tau g_{np}$ within the dilute impurity limit, one expects an intrinsic and a linear scattering time-dependent term from the ${\bm J}^{\rm do}$  component of the thermal current. Surprisingly, we find that the extrinsic part of the ${\bm J}^{\rm do}$ component vanishes identically, $i.e., \ {\bm J}^{\rm do}_{\rm ext} = 0$. The remaining intrinsic part of the ${\bm J}^{\rm do}$ is given by
\be\label{eq:J_do_int}
\bal 
 J^{\rm{do}}_{a,\rm{int}}  &= -\frac{1}{4 } \sum_{n,p, {\bm k}}^{p \ne n} \frac{1}{\e_{np}^2}\m{R}^b_{pn}\m{R}^c_{np} (\e_n + \e_p) (\e_n f^B_{n}- \e_p f^B_{p}) \left(\e_n v_n^{a}- \e_p v_p^{a} \right) E^b_T E^c_T~, \\
 & = - \frac{1}{ 2 } \sum_{n,p,\kb}^{p \ne n} \frac{\m{G}^{bc}_{np}}{\e_{np}^2} \e_n (\e_n + \e_p) \left(\e_n v^n_{a}-\e_p v^p_{a}\right) f^B_n E^b_T E^c_T~. 
\eal 
\ee
Here, we use the same manipulations discussed in Eq.~\eqref{B4}. Similarly, we obtain the nonzero intrinsic and the linear scattering time-dependent extrinsic contributions from the ${\bm J}^{\rm od}$ thermal current component. The intrinsic part of this  current component is determined by the band-resolved quantum metric and has the form, 
\be \label{eq:J_od_int}
J^{\rm{od}}_{a,\rm{int}} = \frac{2}{\hbar} \sum_{n,p,\kb}^{p \ne n} \frac{\m{G}^{ab}_{np}}{\e_{np}} \e_n^2 (\e_n+ \e_p)   \partial_{c} f^B_{n}  E^b_T E^c_T~.
\ee
The extrinsic part of ${\bm J}^{\rm od}$ is governed by the band-resolved Berry curvature. Explicit calculation of this term yields
\be\label{eq:J_od_extrinsic} 
J^{\rm{od}}_{a,\rm{ext}} =  \frac{\tau}{2 \hbar^2} \sum_{n,p,\kb}^{p \ne n} \e^2_n (\e_n+ \e_p) \Omega^{ab}_{np} \partial_c f^B_n E^b_T E^c_T~.
\ee 
Likewise, we extract the intrinsic and extrinsic parts of the thermal current component ${\bm J}^{\rm oo}$. The $\tau$ dependence of this current is determined by the factor $g_{np}$, which is independent of $\tau$ in DIL. Thus, ${\bm J}^{\rm oo}$ has only an intrinsic part. We simplify the intrinsic part  of $J^{\rm oo}_{a}$  as follows:

\bea\label{j_oo_append}
J^{\rm oo}_{a,{\rm int}} &=&  -\frac{1}{4\hbar } \sum_{n,p,\kb}^{p \ne n}   ( \e_n + \e_p ) \m{R}^a_{pn} \bigg[  ( \e_n + \e_p ) \m{D}^b_{np} \left( \frac{1}{\e_{np}}\m{R}^c_{np} \xi_{np} \right) \nn \\ 
&& \ + i \sum_q^{q\neq n \neq p} \left( \frac{1}{\e_{nq}}\m{R}^{c}_{nq}\m{R}^{b}_{qp}({\e}_n +{\e}_q )\xi_{nq} - 
\frac{1}{\e_{qp}} \m{R}^{b}_{nq}\m{R}^{c}_{qp}({\e}_q +  {\e}_p )\xi_{qp} \right)\bigg] E^b_T E^c_T~.
\eea  
Now, focus on the first term of the above equation, which we denote by $J_{I}$ as  
\bea  
J_{I} &=& -\frac{1}{4\hbar } \sum_{n,p,\kb}^{p \ne n}   ( \e_n + \e_p )^2 \m{R}^a_{pn} \m{D}^b_{np} \left( \frac{1}{\e_{np}}\m{R}^c_{np} \xi_{np} \right)E^b_T E^c_T~. \nn \\
&=& -\frac{1}{4\hbar} \sum_{n,p,\kb}^{p\ne n }  ( \e_n + \e_p )^2 \m{R}^a_{pn} \left( \partial_b \left( \frac{1}{\e_{np}}\m{R}^c_{np} \xi_{np} \right) - i ({\m R}^b_{nn} - {\m R}^b_{pp})\left( \frac{1}{\e_{np}}\m{R}^c_{np} \xi_{np} \right) \right) E^b_T E^c_T~.
\eea 
With the help of algebraic manipulations, we can write the first term of the rounded bracket as 
\be   
\m{R}^a_{pn}  \partial_b \left( \frac{1}{\e_{np}}\m{R}^c_{np} \xi_{np} \right) 
=  \partial_b \left( \frac{1}{\e_{np}}  \m{R}^a_{pn} \m{R}^c_{np} \xi_{np} \right) 
 -  \frac{1}{\e_{np}}\m{R}^c_{np} \xi_{np} \partial_{b}\m{R}^a_{pn}
\ee  
In this way, we can modify $J_{I}$ as 
\be 
J_{I} = -\frac{1}{4\hbar} \sum_{n,p,\kb} ( \e_n + \e_p )^2 \left[ \partial_b \left( \frac{1}{\e_{np}}  \m{R}^a_{pn} \m{R}^c_{np} \xi_{np} \right) 
- \frac{1}{\e_{np}}\m{R}^c_{np} \xi_{np} {\m D}^b_{pn} {\m R}^a_{pn} \right]E^b_T E^c_T~.
\ee 
Now, we modify the second part of eq.~\eqref{j_oo_append} by exchanging the dummy indices $i.e., \ q \leftrightarrow p \ and \ q \leftrightarrow n $. Thus, we can write the second part of Eq.~\eqref{j_oo_append} as 
\be 
J_{II} = -\frac{i}{4\hbar} \sum_{n,p,\kb}^{p \ne n } ( \e_n + \e_p ) {\m R}^c_{np} \frac{\xi_{np}}{\e_{np}} \sum_q^{q \ne n \ne p} ( ( \e_n + \e_q ) {\m R}^a_{qn}{\m R}^b_{pq} - ( \e_q + \e_p ){\m R}^a_{pq}{\m R}^b_{qn})E^b_T E^c_T~.
\ee 
With the help of $J_{I}$ and $J_{II}$, the intrinsic current $J^{oo}_{a,int}$ will be reduced to the following expression:
\bea  
J^{oo}_{a,int} &=& -\frac{1}{4\hbar} \sum_{n,p,\kb}^{p \ne n} ( \e_n + \e_p ) \Bigg[ ( \e_n + \e_p ) \partial_b \left( \frac{1}{\e_{np}}  \m{R}^a_{pn} \m{R}^c_{np} \xi_{np} \right) \nn \\ 
&&- \frac{{\m R}^c_{np} \xi_{np}}{\e_{np}}\left( ( \e_n + \e_p ){\m D}^b_{pn} {\m R}^a_{pn} - i \sum_{q \ne (n,p)} ( ( \e_n + \e_q ){\m R}^a_{qn}{\m R}^b_{pq} - ( \e_q + \e_p ){\m R}^a_{pq}{\m R}^b_{qn}) \right) \Bigg]E^b_T E^c_T~.
\eea  
With the help of sum rule~\cite{watanbe_PRX2021_chiral}, we further simplify the above equation by showing that
\be  
{\m D}^a_{pn}{\m R}^b_{pn} - {\m D}^b_{pn}{\m R}^a_{pn} = -i \sum_{q \ne (n,p)} ( {\m R}^a_{qn}{\m R}^b_{pq} - {\m R}^a_{pq}{\m R}^b_{qn}))~.
\ee  
Thus, the intrinsic component of $J^{\rm oo}$ becomes 
\bea\label{j_oo_e8}  
J^{oo}_{a,int} &=& -\frac{1}{4\hbar} \sum_{n,p,\kb}^{p \ne n} ( \e_n + \e_p )^2\left[\partial_b \left( \frac{1}{\e_{np}}  \m{R}^a_{pn} \m{R}^c_{np} \xi_{np} \right)  - \frac{{\m R}^c_{np} \xi_{np}}{\e_{np}} {\m D}^a_{pn} {\m R}^b_{pn}\right]E^b_T E^c_T \nn \\ 
&& - \frac{i}{4\hbar} \sum_{n,p,q,\kb}^{n \ne p \ne q} (\e_n + \e_p) \frac{{\m R}^c_{np} \xi_{np}}{\e_{np}} \left( ( \e_q - \e_p ){\m R}^a_{qn}{\m R}^b_{pq} + ( \e_n - \e_q ){\m R}^a_{pq}{\m R}^b_{qn}\right)E^b_T E^c_T ~. 
\eea  
The second term on the right-hand side of the above equation will be nonzero only for systems having three or more bands. Therefore, we omit this term for a two-band system and focus on the first part of Eq.~\eqref{j_oo_e8}. The permutation symmetry of indices $b$ and $c$ renders the non-trival part of the U(1) covariant derivative ${\m D}^a_{pn} {\m R}^b_{pn}$. This can be viewed by symmetrizing indices $b$ and $c$ followed by the exchange of dummy indices $n \leftrightarrow p$. With this manipulation, we can write the second part in the first term of the above equation as 
\bea 
J_2 &=& \frac{1}{4\hbar} \sum_{n,p,\kb} \frac{( \e_n + \e_p )^2}{\e_{np}} \xi_{np} {\m R}^c_{np} \partial_a {\m R}^b_{pn} E^b_T E^c_T~, \nn \\  
&=& \frac{1}{4\hbar} \sum_{n,p,\kb} \frac{( \e_n + \e_p )^2}{\e_{np}} \e_n f^B_n  \partial_a ({\m R}^c_{np}{\m R}^b_{pn})E^b_T E^c_T~, \nn \\ 
&=& \frac{1}{4\hbar} \sum_{n,p,\kb} \frac{( \e_n + \e_p )^2}{\e_{np}} \e_n f^B_n  \partial_a {\m G}^{bc}_{np} E^b_T E^c_T~.
\eea 
In obtaining this result, we have expanded the $\xi_{np}$ in the second line of the above equation and then used the exchange of dummy indices. Likewise, we transform the first part in the first term of Eq.~\eqref{j_oo_e8} as 
\bea
J_1 &=& -\frac{1}{2\hbar} \sum_{n,p,\kb} ( \e_n + \e_p )^2 \partial_b \left( \frac{\e_n f^B_n }{\e_{np}} {\m G}^{ac}_{np} \right)E^b_T E^c_T~, \nn \\ 
&=& \frac{1}{2\hbar} \sum_{n,p,\kb} \left( \frac{\e_n f^B_n }{\e_{np}} {\m G}^{ac}_{np} \right) \partial_b ( \e_n + \e_p )^2 E^b_T E^c_T~. 
\eea 
In the last step, we have shifted the derivative part to $(\e_n + \e_p)^2$ using integration by part. This is facilitated by the fact that the boundary term is negligible. 
Finally, using all the above manipulations, we write a simplified form of Eq.~\eqref{j_oo_append} for a two-band system as
\be  
J^{\rm oo}_{a,{\rm int}} = \frac{1}{4\hbar} \sum_{n,p,\kb}^{p \ne n}   \frac{( \e_n + \e_p ) \e_n}{\e_{np}} f_n^B \left[ ( \e_n + \e_p ) \partial_a{\m G}^{bc}_{np} + 4{\m G}^{ac}_{np} \partial_b( \e_n + \e_p )\right]E^b_T E^c_T~.
\ee  
%

%
%
%


\section{\label{gauge invariance} Gauge invariance of the band resolved quantum metric (\texorpdfstring{${\m G}_{np}$}{Gnp})}
In this section, we will discuss the gauge invariance of the quantum metric. We have defined quantum metric as ${\m G}^{ab}_{np} = {\rm Re}[ {\m R}^a_{np} {\m R}^{b}_{pn}]$ with ${\m R}^a_{np} = i\bra{u_n}\partial_a\ket{u_p}$ and $n \ne p$. In order to check gauge invariance let's consider that the eigenstates $\ket{u_n}$ and $\ket{u_p}$ transform under gauge transformation as  $\ket{\tilde{u}_n} = e^{i\alpha_n} \ket{u_n}$ and $\ket{\tilde{u}_p} = e^{i\alpha_p} \ket{ u_p}~$, respectively. Through this, we can show
that $\partial_a \ket {\tilde{u}_n } = i \partial_a \alpha_n e^{i \alpha_n} \ket{u_n} + e^{i \alpha_n} \partial_a \ket{u_n}$. So, under gauge transformation, the Berry connection changes as 
\be  
\bal  
{\tilde{\m R}}^a_{np} &= i \langle \tilde{u}_n\vert \partial_a \tilde{u}_p \rangle~, \\
&= i \langle \tilde{u}_n\vert \left[ i \partial_a \alpha_p e^{i \alpha_p} \vert u_p \rangle + e^{i \alpha_p} \partial_a \vert u_p\rangle \right]~, \\
&= e^{i(\alpha_n - \alpha_p)} \left[- \langle u_n \vert u_p \rangle \partial_a\alpha_p + i \langle u_n \vert \partial_a u_p \rangle   \right]~, \\
{\tilde{\m R}}^a_{np}  &=  e^{i(\alpha_n - \alpha_p)} \left[ {\m R}^a_{np} - \partial_a \alpha_p \delta_{np} \right]~.
\eal  
\ee   
Now, we investigate, ``How does the product of two Berry connections transform under gauge transformation?" With the help of the above gauge transformed Berry connection, we can easily show:
\be  
\bal   
{\tilde{\m R}}^a_{np}{\tilde{\m R}}^b_{pn}
&= e^{i(\alpha_n - \alpha_p)} \left[ {\m R}^a_{np} - \partial_a \alpha_p \delta_{np} \right] e^{i(\alpha_p - \alpha_n)} \left[ {\m R}^b_{pn} - \partial_b \alpha_n \delta_{pn} \right]~, \\
&= {\m R}^a_{np} {\m R}^b_{pn} - \delta_{np} \left[ {\m R}^a_{np} \partial_b \alpha_n + {\m R}^b_{pn}\partial_a\alpha_p  - \partial_a \alpha_p \partial_b \alpha_n \right]~.
\eal  
\ee  
Since $n \ne p$, then the second term in the second line of the above equation will vanish due to the Kronecker delta function. Consequently, we have ${\tilde{\m R}}^a_{np}{\tilde{\m R}}^b_{pn} = {\m R}^a_{np}{\m R}^b_{pn}$. It means that the product of the two band-resolved Berry connections is gauge invariant. Likewise, we investigate the gauge transformation of the band-resolved quantum metric, which is written as $
\tilde{\m G}^{ab}_{np} = {\rm Re}[ {\tilde{\m R}}^a_{np}{\tilde{\m R}}^b_{pn}]~.$ As the ${\tilde{\m R}}^a_{np}{\tilde{\m R}}^b_{pn}$ is gauge invariant, so does its real and imaginary parts. Therefore, we conclude that the band resolved quantum metric (${\m G}^{ab}_{np}$) and Berry curvature ($\Omega^{ab}_{np}$) are also gauge invariant quantities.    

\section{Symmetry Analysis}\label{symmetry_and_results}
With zero DMI, the two magnon bands become degenerate at $K, K'$ points.
The inclusion of the second-order Heisenberg interaction and the out-of-plane Zeeman field shifts the energy of the Dirac points rather than creating a gap~\cite{PhysRevB.94.075401} (the corresponding terms are proportional to an identity matrix).
Now, we consider the symmetries in the magnon Hamiltonian and the lowest-order perturbation on top of the ground state.
As honeycomb lattice is a  bipartite lattice, we can write $\bold{S}_{A}=S \hat{z}+\delta\bold{S}_{A}$ and $\bold{S}_{B}=S \hat{z}+\delta\bold{S}_{B}$. The first-order Heisenberg coupling can be written as ($S=1$), 

\begin{equation}\label{magnonHfm}
H_{J}=\sum_{\langle AB \rangle}-J_{1}(1+\delta \bold{S}_{B}^{z}+\delta \bold{S}_{A}^{z}+\delta \bold{S}_{B} \cdot \delta \bold{S}_{A}),
\end{equation}

where, $\langle AB \rangle$ denotes the nearest neigbour sites. As we are interested in the transport properties of the magnons, all the symmetry operator acts on $\delta \bold{S}_{A}$ and $\delta \bold{S}_{B}$ leaving the underlying ferromagnetic ground state unchanged. This Hamiltonian is invariant under a $Tc_{x}$ symmetry, where $T$ is the time reversal operator and $c_{x}$ is a 180$^\circ$ rotation about $\bm{x}$ axis in spin space. Action of the operators is as follows: $c_{x} \delta \bold{S}_{A/B}^{x} : \delta \bold{S}_{A/B}^{x}$, $c_{x} \delta \bold{S}_{A/B}^{y} : -\delta \bold{S}_{A/B}^{y}$, $c_{x} \delta \bold{S}_{A/B}^{z} : -\delta \bold{S}_{A/B}^{z}$ and $T \  \delta \bold{S}_{A/B} : -\delta \bold{S}_{A/B}$. If the magnon Hamiltonian is invariant under this symmetry, then the equation of motion of the magnon wavepacket will also remain invariant. This leads to $\e(\kb)=\e(-\bold{k})$ and $\Omega(\kb)=-\Omega(-\kb)$ ($\e(\kb)$ and $\Omega(\kb)$ are magnon dispersion and Berry curvature respectively). The midpoint of the A-B link in the honeycomb lattice is the inversion center. So, under inversion operation $ I \delta \bold{S}_{A}: \delta \bold{S}_{B}$. From Eq.~\eqref{magnonHfm}, it is clear that $H_{J}$ is also invariant under the inversion symmetry that forces the relation $\Omega(-\bm{k})=\Omega(\kb)$. So, the Berry curvature is identically zero throughout the whole Brillouin zone. Now, when nonzero DMI is turned on, the additional term in the magnon Hamiltonian is written as, 

\begin{equation}
H_{D}=D\sum_{\langle \langle AA' \rangle \rangle}(\delta S_{A}^{x}\delta S_{A'}^{y}-\delta S_{A}^{y}\delta S_{A'}^{x})-(A \longrightarrow B),
\end{equation}

where $\langle \langle AA' \rangle \rangle$ denotes the second nearest neighbor link. This breaks the $Tc_{x}$ symmetry, changes the band topology, and opens up a gap at $K, K'$ points hence a nonzero Berry
curvature, giving rise to a linear magnon Hall effect. However, the dispersion continues to be an even function of the Bloch momentum. As in this work, we want to explore the magnon transport properties coming entirely due to the presence of the quantum metric term, we keep $D=0$, in this case, the linear (depends on the Berry curvature) and nonlinear extrinsic (depends on the Berry curvature dipole) Hall coefficient will remain zero, and we will only have contributions coming from the nonlinear intrinsic (quantum metric dependent) and Drude (quantum metric/Berry curvature independent) part. For the Hamiltonian in Eq.~(\ref {Hamiltonian_model}) with $D=0$,
\begin{equation*}
    \e(\kb)=\e(-\kb ), \ \ G^{ab}_{np}(\kb)=G^{ab}_{np}(-\kb)
\end{equation*}
When we sum over the entire Brillouin zone, the overall contributions of the nonlinear intrinsic current cancel (Eq.~\eqref{eq:tot_th_cur_int}) out since the integrand is an odd function of momentum $\kb$.
In order to explicitly break the valley symmetry, we include the term $B_{s}(k_{x}+k_{y})\sigma^{0}$ in the Hamiltonian. This invalidates the relation $\e(\kb)=\e(-\kb)$ keeping the eigenfunction hence the quantum metric unchanged. This term can be engineered for graphene systems by applying strain and in-plane electric fields~\cite{PhysRevB.78.045415, Goerbig}. Before going into the details on how to engineer such a term in ferromagnetic insulators, let us go back and discuss the presence of the $J_{2}$ term in our original Hamiltonian. For $D=0$, the contributions coming from $J_{2}$ term go into the $\sigma^{0}$, but as it comes with $\cos{} (\cdots)$ terms, and its effect on low energy model near $K, K'$ will be proportional to $\kb^2$ which will not break the valley symmetry. Even the anisotropic $J_{2}$ results in the tilting of Dirac cones with $\kb$ linear contributions near Dirac points, but does not break the $K, K'$ symmetry. The presence of a substrate  has been reported to break this symmetry by opening up a gap at $K, K'$ points nonsymmetrically~\cite{PhysRevB.101.205425}.

\subsubsection*{Aharonov-Casher effect and the valley symmetry breaking term}


In the presence of an external electric field and spin-orbit coupling, the Heisenberg Hamiltonian is modified by the Aharonov-Casher (AC) phase~\cite{ACphase1,ACphase3} and its given by\cite{ACphase2},

\be
\sum_{i,j}J_{ij}[S^{z}_{i}S^{z}_{j}+\dfrac{1}{2}(e^{2i\chi_{ij}}S^{+}_{i}S^{-}_{j}+e^{-2i\chi_{ij}}S^{-}_{i}S^{+}_{j})]~.
\ee
Here, the $\chi_{ij}$ phase depends on the magnitude of the spin-orbit interaction and the external electric field and also on the distance between the site $i$ and $j$. 
If we tune our external parameter in such way that $\chi_{ij} \approx \pi/4$ for the second nearest neighbor interaction then effectively the Hamiltonian will have terms like $iS^{+}_{i}S^{-}_{j}-iS^{-}_{i}S^{+}_{j}$ for the second nearest neighbor. In the magnon Hamiltonian this will induce a term that is proportional to $\sin(k_{x}+k_{y}) \sigma_{0}$. Here we note two important points, firstly the AC phase $\chi_{ij}$ is proportional to the distance between neighboring sites (terms for the AA and BB links will be exactly equal)  so it will not generate a term that is proportional to $\sigma_{z}$, as a consequence, it will never gap out the magnon spectrum at $K, K'$ points; secondly, the AC phase will induce an effective DMI coupling~\cite{ACphase2} between the nearest neighbor spins but it will go in the off-diagonal terms in the Hamiltonian (terms with $\sigma_{x}$ and $\sigma_{y}$) and will not alter the topology of the system. In this context, we would also like to mention that recently there have been few works that have explored the possibilities of electric field control of magnons in two-dimensional heterostructures~\cite{ACphase4,ACphase5}. 

 {To estimate the magnitude of the effective $B_s$ induced by the  AC phase, we note that for a small value of the quantum phase, $B_s \propto \chi_{ij}$. In a Mach-Zehnder spin-wave interferometer, the quantum AC phase $\Phi_{AC}$ is the sum of the quantum phase due to the AC effect, which a flipping spin acquires when it runs around the ring once, and it is given by~\cite{ACphase3} $N \chi_{ij}/ 2\pi$, where $N$ is the total number of sites. If we assume the distance between the two nearby sites to be of the order of $a= 1$ \AA, and the radius ($r_0$) of the interferrometer~\cite{ACphase2} to be approximately 50 $nm$, then the number of sites $N=2\pi r_0/a \approx 3 \times 10^3$. With a reasonable radial electric field value of $0.1$ V/\AA, the corresponding quantum AC phase is given by,
\be
\Phi_{AC}=\dfrac{\mu E r_0}{\hbar c^2} \approx 10^{-4}~.
\ee
Here, $\mu$ is the magnetic moment in units of Bohr magneton. This corresponds to $B_s \approx 10^{-6}$. For this $B_s$ value, we estimate the corresponding intrinsic nonlinear thermal Hall current to be of the order of $10^{-17}$ W/m, which is in the experimentally  measurable range.}

\twocolumngrid
\bibliography{References}


\end{document}